\pdfoutput=1

\documentclass{emulateapj}

\newcommand{\dasec}{\hbox{$.\!\!^{\prime\prime}$}}     
\newcommand{\damin}{\hbox{$.\!\!^{\prime}$}}           
\newcommand{\sdeg}{$^{\circ}$}                         
\newcommand{\ddeg}{\hbox{$.\!\!^\circ$}}               
\newcommand{\kms}{km s$^{-1}$}
\newcommand{\hii}{H\ {\small II}}                      

\begin{document}


\title{GBT Multiwavelength Survey of the Galactic Center Region}
\author{C. J. Law\altaffilmark{1,2}, F. Yusef-Zadeh\altaffilmark{1}, W. D. Cotton\altaffilmark{3}, and R. J. Maddalena\altaffilmark{4}}
\altaffiltext{1}{Department of Physics and Astronomy, Northwestern University, 2145 Sheriadan Road, Evanston, 60208 IL, USA}
\altaffiltext{2}{Astronomical Institute ``Anton Pannekoek'', University of Amsterdam, Kruislaan 403, 1098 SJ Amsterdam, Netherlands}
\altaffiltext{3}{National Radio Astronomy Observatory, 520 Edgemont Road, Charlottesville, VA 22903, USA}
\altaffiltext{4}{National Radio Astronomy Observatory, P.O. Box 2, Rt. 28/92, Green Bank, WV 24944, USA}

\begin{abstract}
We describe the results of a radio continuum survey of the central 4\sdeg$\times$1\sdeg\ with the 100 m Green Bank Telescope (GBT) at wavelengths of 3.5, 6, 20, and 90 cm.  The 3.5 and 6 cm surveys are the most sensitive and highest resolution single dish surveys made of the central degrees of our Galaxy.  We present catalogs of compact and extended sources in the central four degrees of our Galaxy, including detailed spectral index studies of all sources.  The analysis covers star-forming regions such as Sgr B and Sgr C where we find evidence of a mixture of thermal and nonthermal emission.  The analysis quantifies the relative contribution of thermal and nonthermal processes to the radio continuum flux density toward the GC region.  In the central 4\sdeg$\times$1\sdeg\ of the GC, the thermal and nonthermal flux fractions for all compact and diffuse sources are 28\%/72\% at 3.5 cm and 19\%/81\% at 6 cm.  The total flux densities from these sources are $783\pm52$ Jy and $1063\pm93$ Jy at 3.5 and 6 cm, respectively, excluding the contribution of Galactic synchrotron emission.
\end{abstract}


\section{Introduction}
The central few hundred parsecs of the Milky Way comprise a region in the Galaxy unique for its high stellar density, intense ionizing radiation field, massive black hole, enhanced density of cosmic rays, and unusual magnetized structures \citep{f04,r01,y07,b87,y04}.  The extent of the GC region is roughly 400 pc in diameter, defined by a region with relatively high gas density \citep[$n_{\rm{H_2}}\gtrsim10^4$ cm$^{-3}$;][]{h98}.  This region, sometimes called the Galactic nucleus or ``central molecular zone'', produces 5\%--10\% of the Galaxy's infrared and Lyman continuum luminosity and contains 10\% of its molecular gas \citep{b87,m96}.  The density and physical diversity of objects in the GC region make it highly complex and require a wide range of observations to unravel.

At the simplest level, it is important to understand how the basic components of the region interact.  Nonthermal emission, in the form of supernova remnants (SNRs) and nonthermal radio filaments \citep[NRFs;][]{y84}, dominates the cm-wavelength emission in the region \citep[e.g.,][]{a79,d95}.  The relativistic component of the ISM has also been observed at TeV energies as a diffuse source tracing the molecular gas distribution \citep{a06}.  However, it is not clear how these energetic electrons traced by the nonthermal emission affect other components of the GC interstellar medium.  For example, energetic electrons may explain the unusually high ambient gas temperature in the region and subsequent low star formation efficiency \citep{y07}.  Thermal emission traces star formation regions and photoionized clouds.  An accurate census of thermal emission can constrain the amount of star formation occurring there or find new star forming regions.  Separating the thermal and nonthermal processes also allows an estimate of their basic properties \citep[$n_e$, $B$;][]{h96}.

Single-dish radio continuum observations are a useful tool for studying the large-scale properties of the GC region.  Several other single-dish surveys of radio continuum emission from the GC region have been conducted \citep{a79,ha87,r90,h92,d95}.  \citet{a79} observed the Galactic disk from $l=60$\sdeg\ to the GC region near 6 cm with the 100 m Effelsberg telescope.  This was one of the first single-dish surveys of the GC region with a resolution of a few (2\damin6) arcminutes and it discovered many compact and diffuse sources.  Since that time, the Parkes 64 m and Effelsberg 100 m telescopes have been used to create complete surveys of the Galactic disk in the northern and southern celestial skies near 12 cm, which were sensitive to faint emission on large scales \citep{r90,d95}.  These studies revealed dozens of new SNR candidates, compact \hii\ regions, and Galactic loops and spurs, vividly demonstrating the chaotic structure of the Galactic interstellar medium.  

Although most of the world's best radio telescopes have surveyed the radio continuum in the GC region, there has been limited study of spectral indices on arcminute scales.  We were motivated to extend upon previous observations using the largest, fully-steerable telescope, the Green Bank Telescope\footnote{The GBT is operated by the National Radio Astronomy Observatory, which is a facility of the National Science Foundation operated under cooperative agreement by Associated Universities, Inc.} (GBT).  The high resolution and sensitivity of the GBT observations allow us to separate and quantify flux from all thermal and nonthermal emitters.  Section \ref{gcsurvey_obs} describes the observations and data reduction.  In \S\ \ref{gcsurvey_res}, the results of the survey are described, including the compilation of compact and extended source catalogs at 3.5 and 6 cm and a detailed discussion of the radio continuum properties for each source in the region.  We calculate the percentage of flux from sources in the central degrees of the Galaxy from thermal/nonthermal processes at 3.5 and 6 cm.  Finally, \S\ \ref{gcsurvey_con} summarizes the results of this analysis.  A study of the continuum emission from the GC lobe \citep{s84} is not included here, but will be discussed in detail in a later paper \citep{l08}.

\section{Observations and Data Reductions}
\label{gcsurvey_obs} 
\subsection{Description of Observations}
In June 2003, we surveyed the central degrees of the Galaxy with the GBT at 3.5, 6, 20, and 90 cm.  The GBT is located in Green Bank, West Virginia at a latitude of 38\sdeg26\arcmin\ N.  The GBT is unique in that it has a large (100 m by 110 m) unblocked aperature that is fully steerable, with an elevation range of 5--90\sdeg.   Observations were conducted over five days in August of 2002 for a total observing time of 30 hours.  At 3.5, 6, and 20 cm, observations were made with the Digital Continuum Receiver in ``on-the-fly'' mapping mode, while at 90 cm the observations were made with the Spectral Processor.  After flagging the 90 cm data for radio frequency interference, the data were averaged in frequency and treated identically to the higher frequency data.  Observations had bandwidths of 320, 320, 20, and 40 MHz at 3.5, 6, 20, and 90 cm.

\begin{figure}[tbp]
\plotone{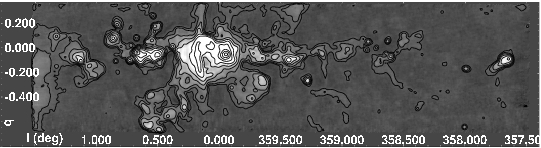}
\plotone{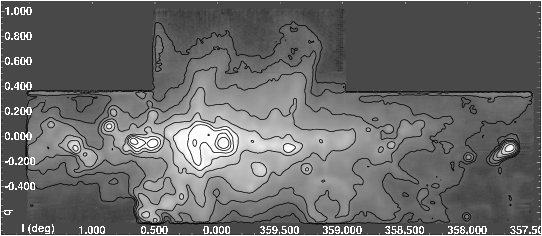}
\caption{The GBT radio continuum survey of GC region at 3.5 and 6 cm is shown in the top and bottom panels, respectively.  The surveys cover a similar region of roughly 4\sdeg$\times$1\sdeg, although the 6 cm survey also covered a region at higher positive latitudes.  Contours for the 3.5 cm survey are at levels of $0.05*2^n$ Jy beam$^{-1}$, for $n=0-9$ and a beam size of 88\arcsec.  The contours for the 6 cm survey are identical, but for $n=1-9$ and a beam size of 153\arcsec.  Galactic coordinates are shown on each image.  \label{mapscx}}
\end{figure}

\begin{figure}[tbp]
\includegraphics[width=0.65\textwidth]{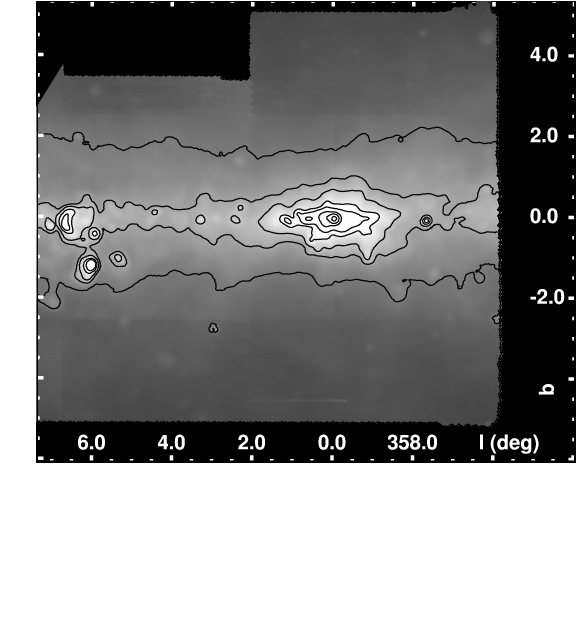}

\includegraphics[width=0.65\textwidth]{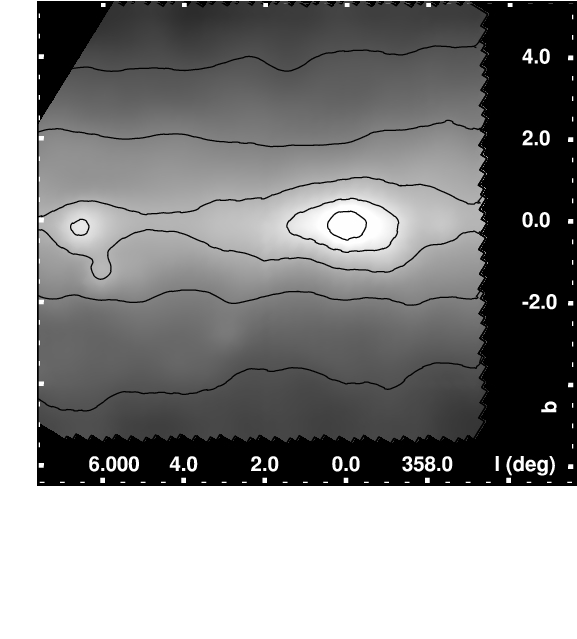}
\caption{The GBT radio continuum survey of GC region at 20 and 90 cm is shown in the top and bottom panels, respectively.  The surveys cover a similar region of roughly 10\sdeg$\times$10\sdeg.  Contours for the 20 cm survey are at 5, 10, 15, 20, 30, 40, 80, 160, and 320 Jy beam $^{-1}$, with a beam size of 9\arcmin.  Contours for the 90 cm survey are at 80, 160, 320, 640, and 1280 Jy beam $^{-1}$, with a beam size of $38\damin8$.  \label{mapspl}}
\end{figure}

All observations surveyed at least a 4\sdeg$\times$1\sdeg\ area roughly centered on the GC, which includes the molecular zone in the central 3\sdeg\ ($\sim$400 pc);  the 6, 20, and 90 cm maps also surveyed beyond this region.  The 3.5 and 6 cm surveys, shown in Figure \ref{mapscx}, have similar coverage, while the 20 and 90 cm surveys, shown in Figure \ref{mapspl}, cover the central ten degrees.  The spatial coverage, resolution, and typical $1\sigma$ sensitivity of these maps are shown in Table \ref{mapstats}.

Flux calibration was done by adding a calibrated noise source to every other integration.  The brightness of the noise source is estimated from observations of 3C286 and gives a conservative absolute flux accuracy of 5\%.  The final amplitude calibration is made from the median brightness of the noise source for each scan.

No simultaneous measurement of the sky brightness was made during these observations, so a few techniques were used to estimate the noncelestial background for each map.  First, an initial esimate of the sky brightness is done by observing a position far from the Galactic plane, near 3C286.  Second, the atmospheric opacity and temperature are estimated from weather monitoring stations and subtracted from the data.  Finally, an iterative scheme was used to separate the constant celestial brightness distribution from the time variable atmospheric and spillover contributions.  An image of the sky was made with the initial estimates of the atmospheric and receiver contributions removed from the data.  This model sky brightness was then subtracted from the data and a high pass time filtering produced a refined estimation of the atmospheric and spillover contributions.  This refined calibration was then used to correct the data, resulting in an improved estimate of the sky distribution.  Several iteration with decreasing time constants of the high pass filtering yielded the results presented.  It is important to note that this technique cannot distinguish between slow variations in the atmospheric brightness and true celestial changes, which effectively introduces a local ``zero point''.  Thus, measurements of brightnesses and flux densities require the subtraction of a local background, especially in the 3.5 and 6 cm maps, where atmospheric effects are stronger.

Each of the radio continuum maps was observed with orthogonal, ``basket-weaving'' patterns in Galactic coordinates.  The redundancy of observing each position in the map in two orthogonal scans is used to identify the flux contribution from noncelestial sources.  The final images were made by convolving the data with a Gaussian kernel and resampling onto the image grid.  Maps were made with ``Obit'', a group of software packages designed to handle single-dish and interferometric radio astronomy data, such as AIPS data disks or FITS files.  ``ObitSD'' is a low-level addition to the software package and is designed for making maps from on-the-fly data.

The 3.5 and 6 cm surveys have sensitivities of 8 and 10 mJy beam$^{-1}$, which makes them one of the highest-resolution and most sensitive single-dish surveys of the region made at these wavelengths.  At wavelengths near 3.5 cm, the GC region has previously been best surveyed with 45 m Nobeyama Radio Observatory and 64 m Parkes Telescope to sensitivities of 15 and 30 mJy beam$^{-1}$, respectively, with resolutions of about 2\damin5 \citep{ha87,h92};  the present survey surpasses this previous work in resolution and sensitivity.  The 6 cm survey of \citet{a79} used the 100 m Effelsberg telescope, which has a comparable resolution of 2\damin6 and a sensitivity of 5 mJy beam$^{-1}$, or about twice as sensitive as the present work.  An advantage of the present work is that it was planned as a multiwavelength campaign, so the observations at 3.5, 6, 20, and 90 cm were calibrated by similar techniques;  this makes the survey robust and especially suited to studying spectral indices.

\subsection{Basic Source Analysis}
Compact sources were identified by eye by searching for sources not dominated by extended emission.  Source properties were measured by JMFIT of AIPS.  JMFIT fits a 2-D Gaussian to an image, assuming an initial size and shape equal to the beam size.  The JMFIT routine also fits for the absolute background flux and its first spatial gradient, which is particularly useful for the 6 cm maps.  The output of the source detection is a set of positions, flux densities, sizes for a best-fit Gaussian, plus a residual image.  The residual images were inspected as a test of the fit quality and were generally less than about 10\% of the source peak brightness.  Other source-fitting algorithms, including the IDL astrolib photometry programs and MIRIAD's sfind, were tested on our images, but neither of these produced as trustworthy results as JMFIT.  In particular, the algorithms did not easily produce source-subtracted residual images to enable visual inspection of the fit quality.

All sources detected at 6 and 3.5 cm have a spectral index calculated.  The spectral index is calculated from the integrated flux densities assuming $S_\nu \propto \nu^{\alpha}$, which gives:

\begin{equation}
\label{alpha2}
\alpha_{12} = \rm{log}(S_1/S_2)/\rm{log}(\nu_1/\nu_2)
\end{equation}

\begin{equation}
\label{alpha3}
\sigma_{\alpha_{12}} = (1/\rm{log}(\nu_1/\nu_2))*\sqrt{(\sigma_1/S_1)^2 + (\sigma_2/S_2)^2}
\end{equation}

\noindent where ``1'' and ``2'' refer to the observing frequencies, which are abbreviated as X for 3.5 cm, C for 6 cm, L for 20 cm, and P for 90 cm.  The images were not convolved to the same resolution for measuring the compact source flux densities, since tests with JMFIT show that the integrated flux density of the 3.5 cm image is the same at its natural resolution and after being convolved to match the 6 cm GBT resolution.

Extended sources in the GC region, such as \hii\ regions, SNRs, and nonthermal radio filaments (NRFs), dominate the radio brightness of the GC region.  For a more detailed study of these sources, flux density slices were taken at different positions and orientations through extended sources in order to measure brightnesses and the corresponding spectral index.  Slices were taken from two images convolved to the same resolution;  the convolution size was 2\damin5 for the 3.5/6 cm slices and 9\arcmin\ for the 6/20 cm slices.  To estimate the spectral index of a source, a background was subtracted prior to taking the ratio of the source brightnesses.  This is done by a reduced chi squared fit of a line to a portion of the slice data.  All structure in the slice that does not look like a background was ignored in fitting the background.  The background was fit to each slice independently, although the same background region was used at both frequencies.  The best-fit line was then subtracted from the data prior to calculation of the spectral index.

\begin{figure}[tbp]
\includegraphics[width=0.8\textwidth]{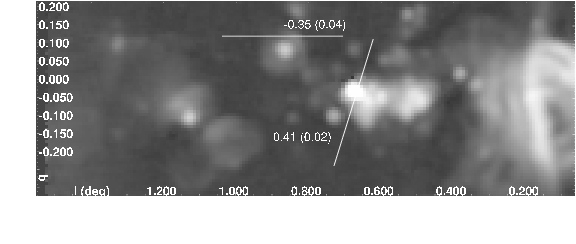}

\includegraphics[width=0.8\textwidth]{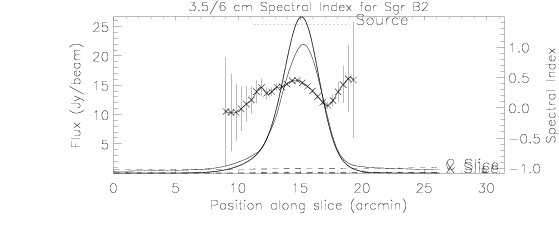}

\includegraphics[width=0.8\textwidth]{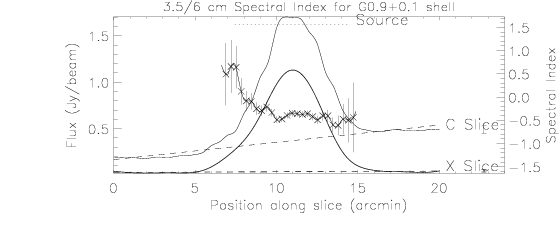}
\caption{\emph{Top}: 3.5 cm image with the locations of two slices used to demonstrate the spectral index slice analysis method.  The slices pass through Sgr B2 and G0.9+0.1, two objects with well-known radio spectral indices.  Both slices are labeled with the 6/3.5 cm spectral index at the brightest part of the slice at 3.5 cm.  \emph{Middle and bottom}:  The flux and spectral index for the slices shown above.  Each plot shows slices for two frequencies with the 3.5 cm (``X'') slice below the 6 cm (``C'') slice.  The x axis shows the entire length of the slice in units of arcmin.  The slice brightnesses are indicated by the left axis and the spectral index value is shown on the right axis.  The best-fit background to each slice is shown as a dashed line.  The dotted line shows the ``source'' region, which is ignored in the determination of the background.  The rms deviation of the background data about the best-fit line is shown as an error bar to the right of each slice.  Parts of the slice with spectral index error, $\sigma_\alpha<1$, are plotted.  \label{slicetest}}
\end{figure}

Figure \ref{slicetest} demonstrates the slice analysis technique on two sources with relatively simple morphologies.  For each slice shown in the image, the flux and spectral index are shown in a plot.  The value of the spectral index shown in the image is measured at the peak brightness of the shorter wavelength image (3.5 cm for 6/3.5 cm comparison and 6 cm for 20/6 cm comparison).  The dashed lines show the best-fit background for the slices, made by ignoring the source emission, which is shown with a dotted line;  sometimes the ``source'' includes parts of the slice that are confused with other sources.  The spectral index measurements were considered trustworthy only if they were found not to vary significantly with $\sim10$\% variations in the background region extent.  The rms deviation of the background data relative to the best-fit line is used as an estimate of the uncertainty in the background-subtracted flux density.  The flux density for each position on the slice and its error are used to calculate the spectral index according the standard relations given in Equations \ref{alpha2} and \ref{alpha3}.  Note that the errors do not include systematic error.  Including the 5\% uncertainty in the flux scale at 3.5 and 6 cm, the systematic spectral index uncertainty is 0.13;  this error should be added in quadrature to all statistical errors quoted in this paper.  The statistical errors in the slice analysis will also underestimate the true error due to small-scale changes in the non-celestial background, particularly at 3.5 cm.  The magnitude of this error underestimate is not clear, but is expected to be insignificant for all but the faintest sources or sources within a few beamwidths of the survey edge.

One slice in Figure \ref{slicetest} covers the brightest part of Sgr B2, which is optically thick and is expected to have a positive (``inverted'') spectral index at cm wavelengths \citep{ga95}.  The plot shows how the background is fit to the parts of the slice with no emission and, after being subtracted from the slice, finds $\alpha_{CX}=0.41\pm0.02$, consistent with expectations for a optically-thick, thermal source \citep{d05}.  The second slice shown in the figure measures the spectral index of the G0.9+0.1 SNR shell.  The slice analysis shows that the source has a nonthermal spectral index of $\alpha_{CX}=-0.35\pm0.04$, as had been previously observed \citep{h87}.  Note that we refer to the spectral index at the peak flux as the source's spectral index, but the spectral index is calculated for every point along the slice.  In some extended sources, the spectral index changes away from the peak flux, and the spectral index distribution is discussed in more detail.

\section{Results}
\label{gcsurvey_res}

\subsection{Compact Source Catalog}
\label{compactsrcsec}
Here we describe the properties of all compact sources found in the 3.5 and 6 cm maps.  A source is considered to be compact if its best-fit width is less than approximately twice the beam FWHM.  Only one of the 36 sources in the 3.5 cm compact catalog has a Gaussian width approaching roughly twice that of the beam (roughly 3\arcmin).  Any source that is not compact in the highest-resolution image (3.5 cm) is considered ``extended'' and described in \S\ \ref{diffsrcsec} and \S\ \ref{slicediff}.

The catalog of compact sources in the 3.5 and 6 cm images are shown in Tables \ref{srcX} and \ref{srcC}, respectively.  In these tables, columns (2)-(5) give the position of the source, columns (6)-(9) give the source peak brightness and its error, the source flux density and its error, and columns (10)-(12) give the major axis, minor axis, and position angle of the best-fit Gaussian to the source.  For comparison, column (13) shows the flux density given in \citet{h92} for sources that are coincident with sources in the present catalog.  The 20 and 90 cm images do not have any distinctly compact sources in the region covered by the 3.5 and 6 cm surveys, so no compact source catalog is given here for those images.  Table \ref{psspix} shows the spectral index for all compact sources detected at 3.5 and 6 cm.  

\citet{h92} used the Parkes 64 m telescope to survey the GC region at 3.5 cm and produce a point source catalog.  Those data have a resolution and sensitivity roughly two to three times poorer than the present observations (2\damin8 beam and 30 mJy beam$^{-1}$, respectively).  Sources in that catalog that are coinicident with the present 3.5 cm survey are shown in Table \ref{srcX} for comparison.  The integrated flux densities observed here are systematically 24\% less than in \citet{h92}, with the greatest difference seen in sources in dense regions such as Sgr B and Sgr C.  Since this difference is more than expected from the calibration uncertainty (6\% for their work and 5\% for the present work), we repeated our source detection on an image convolved to the resolution of \citet{h92}.  The integrated fluxes measured in our convolved map is consistent with that of \citet{h92}, which suggests that the larger beam tends to include more extended emission in the integrated flux of compact sources.  This caveat should be considered when using compact source fluxes, especially in regions with lots of extended emission.

The 6 cm catalog in Table \ref{srcC} was compared to catalogs in the literature to confirm those results, but no single-dish survey at this wavelength could be found.  In the VLA study of Sgr E by \citet{g93}, two sources (J174203--300405 and J174227--295559) are associated with relatively unconfused sources in the present 6 cm catalog.  Also, in the 6 cm VLA survey of \citet{b94}, three sources can be compared to the GBT survey.  In all these cases, the VLA flux densities range from 2--70\% of that measured by the GBT 6 cm survey;  there is likely to be flux missing from the interferometric observations.

To aid interpretation of the source catalogs, we estimated the number of extragalactic sources expected in the survey region.  \citet{h05} fit a seventh-order polynomial to the flux distribution of 20 cm compact sources from several high Galactic latitude catalogs.  We estimate the number of extragalactic sources by integrating over this distribution and assuming a spectral index of --0.7.  At 3.5 and 6 cm, we estimate 0.4 and 0.2 sources, respectively, in the 4\sdeg$\times$1\sdeg\ survey region with a flux density higher than the faintest compact source detected at each wavelength (0.12 Jy and 0.29 Jy, respectively).

\subsection{Extended Sources}
\label{diffsrcsec}
The following section describes the cataloging and analysis of extended sources observed in the survey region.  The spectral index is studied by integrated fluxes and slices of the maps.  Section \ref{srcex} describes how sources are identified and the integrated spectral index between 3.5 and 6 cm, while \S\ \ref{slicediff} describes the results of slice analysis.

\subsubsection{Extended Source Catalog}
\label{srcex}
Here we describe the construction of a catalog of the extended sources.  All sources not considered compact (i.e., not satisfying $\theta_{\rm{3.5 cm}} < 2*\rm{FWHM}$) were included in the extended source catalog shown in Table \ref{diffsrc}.

\begin{figure}[tbp]
\plotone{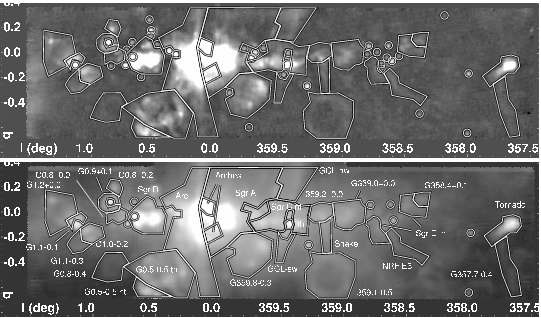}
\caption{\emph{Top}: GBT map of 3.5 cm continuum emission showing the extended sources listed in Table \ref{diffsrc} and the compact sources given in Table \ref{srcX}.  The brightness of the image is shown in a logarithmic scale from 0 to 5 Jy beam$^{-1}$.  \emph{Bottom}:  GBT map of 6 cm continuum emission with the same regions overlaid, except the circles show 6 cm compact sources given in Table \ref{srcC}.  The brightness of the image is shown in a logarithmic scale from 0 to 10 Jy beam$^{-1}$.  The polygon regions define extent of extended sources and are identified with their common names.  The flux density, location, and other properties of the extended sources are given in Tables \ref{diffsrc} and \ref{diffsrcspix}.  \label{diffreg}}
\end{figure}

Figure \ref{diffreg} shows the 3.5 and 6 cm images with extended and compact regions overlaid.  Regions were defined to enclose all flux from an object (in both the 3.5 and 6 cm images) that is believed to have a similar origin.  Often this is simple, such as for the supernova remnant G359.1--0.5, which has a distinctive ring-like shape and has been extensively studied \citep{u92,y95}.  Otherwise, the regions were simply defined by whether it had a thermal or nonthermal spectral index between 6 and 3.5 cm (see \S\ \ref{slicediff}).  In a few cases, the extended sources in Table \ref{diffsrc} have names appended with ``th'' or ``nt'', according to whether the region contains the thermally- or nonthermally-emitting parts of the complex.  The source regions for Sgr A and Sgr B include other sources, so the background-subtracted source flux also subtracts these other contributions, as noted in the table.

Properties of extended sources in the 3.5 and 6 cm images are given in Table \ref{diffsrc}.  Columns (1)-(4) give the commonly-used source name, position, and effective radius, calculated from the area ($R_{eff}=\sqrt{A/\pi}$).   Columns (5)-(8) give the 3.5 cm raw source flux density, the rms uncertainty measured in a background region, and the background-subtracted flux density and its error.  Columns (9)-(12) give the same quantities for the sources at 6 cm.  The source brightness is integrated over all pixel values and then scaled by the ratio of the pixel area to the beam area to get a flux density in Jy.  The scale factor is $(30\arcsec^2)/(1.1331*FWHM^2)$ (as used in AIPS), with $FWHM=88$\arcsec\ and 153\arcsec\ at 3.5 and 6 cm, respectively.  The background flux density is measured over a nearby region.  The rms in the background is scaled to the source area to estimate the uncertainty in the integrated source flux density.  Often this method overestimates the error in the flux density, since the rms in the background is dominated by other sources or Galactic emission.  With the flux density measured at 3.5 and 6 cm, the spectral index and its statistical error is calculated and shown in Table \ref{diffsrcspix}.  The sixth column shows our conclusion on the nature of the radio continuum emission (thermal vs. nonthermal), based on the integrated and slice spectral index analysis.

The spectral index values shown in Table \ref{diffsrcspix} are generally consistent with the spectral indices measured from slices of the data presented in \S\ \ref{slicediff}.  There is a trend for the integrated spectral index values to be lower than the more precise slice analysis.  The integrated spectral index values for G359.2+0.0, G359.1--0.2 (``the Snake''), GCL-NW, G359.8--0.3, and G0.9+0.1 are all at the low end of 1$\sigma$ errors from the slice analysis.  This may be caused by imperfect subtraction of Galactic synchrotron emission, which is stronger at 6 cm than at 3.5 cm, and stronger near the Galactic plane.  Generally, the slice analysis gives a better estimate of the local background, so it gives a more robust estimate of the spectral index.  However, the slice analysis is limited because it cannot measure the integrated flux from an extended source.

\subsubsection{Known and Candidate Supernova Remnants}
The extended source catalog includes eight sources that have previously been identified as known or candidate SNRs.  The types of SNRs represented in the GC region vary from traditional shell-types to those having pulsar-wind neblae to mixed-mophology types.  Table \ref{snrsrc} summarizes the observed characteristics of all supernova remnants in the region.  The results presented here are generally consistent with the high-resolution, 843 MHz observations presented in \citet{g94} and associated papers.  Between 3.5 and 6 cm, the spectral indices range from flat to $\alpha\approx-2$.  A detailed discussion of the sources and comparison to previous work is given in \S\ \ref{slicediff}.

\subsubsection{Thermal/Nonthermal Flux Fractions}
\label{fluxfracsec}
Each extended source in Table \ref{diffsrcspix} has been classified as either thermal or nonthermal according to the spectral index analysis (see \S\ \ref{slicediff}), which allows the fraction of total flux contributed by these processes to be calculated.  Table \ref{fluxfrac} shows the sum of the flux densities for each extended source, with thermal to nonthermal flux ratios of 24\%/76\% at 3.5 cm and 15\%/85\% at 6 cm, with uncertainties of about 4\% in both bands.  The flux density of compact sources with measured spectral indices also contribute significantly, with 39$\pm$0.4 Jy detected at 3.5 cm and 51$\pm$4 Jy at 6 cm (thermal/nonthermal fraction of 43\%/57\%$\pm$2\%;  see Tables \ref{srcX} and \ref{srcC}).  Furthermore, there is about 46$\pm$1 Jy of flux density in sources that are compact at 3.5 cm, but are confused with extended emission at 6 cm (and are thus not in the 6 cm compact source catalog).  Assuming that these other compact sources have a similar distribution of spectral indices, then we can estimate their contribution to the total 3.5 and 6 cm emission toward the GC region\footnote{This assumption should not affect the final values much, since the flux from these sources is smaller than the uncertainty in the total flux}.  Table \ref{fluxfrac} shows the best estimate of the thermal and nonthermal flux fractions in the survey is 28\%/72\% at 3.5 cm and 19\%/81\% at 6 cm.  The total flux densities for all compact and extended sources are $783\pm52$ Jy and $1063\pm93$ Jy at 3.5 and 6 cm, respectively.  The survey region covers the central 4\sdeg$\times$1\sdeg\, equivalent to a physical size of 560 pc$\times$140 pc at a distance of 8.5 kpc.

The first caveat in the interpretation of these results is that these flux fractions are measured for discrete sources and does not consider the diffuse Galactic emission.  Overall, the Galactic emission is dominated by extended synchrotron emission that fills the field observed here, particularly at low frequencies.  Second, the distinction of ``thermal'' and ``nonthermal'' sources by the measured spectral index may not be accurate for sources with a mixture of thermal and nonthermal processes.  For example, it would not be accurate to describe a supernova remnant embedded in a star forming complex as one of these two categories.  While the source regions used to make Table \ref{diffsrc} were defined to separate sources with different structure and emission mechanisms as best as possible, there may be weak emission that is miscategorized by this analysis.

A previous study of the thermal/nonthermal flux distribution toward the GC region was done with the Effelsberg 100 m telescope \citep{s78}.  That work found that the central 350 pc ($2\ddeg5$) had roughly equal contributions from thermal and nonthermal emission at 6 cm, with a total flux density of about 2000 Jy \citep{k83,l94}.  The total flux density is twice that observed here for a similar region, so that work probably included the extended background emission that this work excluded (the work is only available in a thesis).  \citet{me96} found a roughly equal fraction of thermal and nonthermal emission at 6 cm in the central 400 pc$\times$350 pc, with a free-free flux density of $\sim580$ Jy at 6 cm.  The present work finds a similar total flux as that of \citet{me96}, but we identify more of the emission from discrete sources as nonthermal.

A number of recent low-frequency radio continuum, X-ray, and molecular line observations suggest an increase in the cosmic ray ionization rate in the nuclear disk \citep{y06,o05}.  The high fraction of nonthermal emission from the central region shown here is qualitatively consistent with other measurements.

\subsection{Slices of Extended Sources in GBT Images}
\label{slicediff}
\begin{figure}[tbp]
\plotone{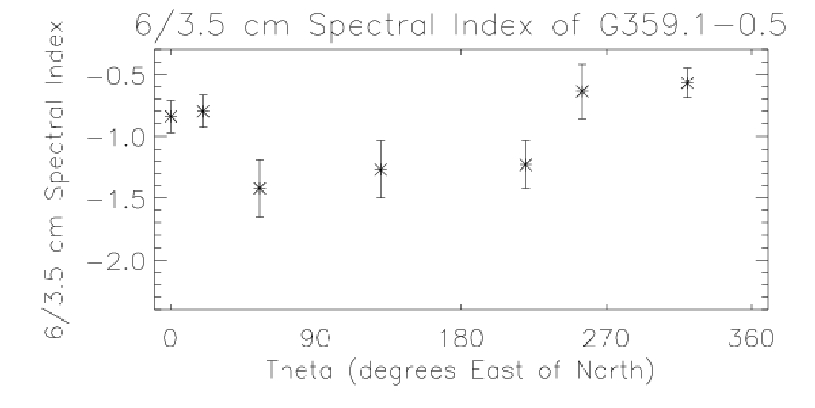}
\caption{Plot of the 6/3.5 cm spectral index values as a function of position on the circular supernova remnant, G359.1--0.5.  Position is represented by an angle measured in degrees, increasing counterclockwise starting at galactic north. \label{plsp}}
\end{figure}

Figures \ref{sptornado} through \ref{spsgrbcl} show details of the 3.5, 6, and 20 cm surveys and the slices used to study the spectral indices of objects between these frequencies.  The figures are ordered with increasing Galactic longitude, starting from the western edge of the 3.5 and 6 cm surveys.  Each figure shows two images convolved to the same resolution with slice positions overlaid.  For each figure, one representative slice is plotted below the images.

\subsubsection{G357.7--0.1 (The Tornado) and G357.7+0.3}
\begin{figure}[tbp]
\includegraphics[bb=0 200 600 600,clip,width=6.5in]{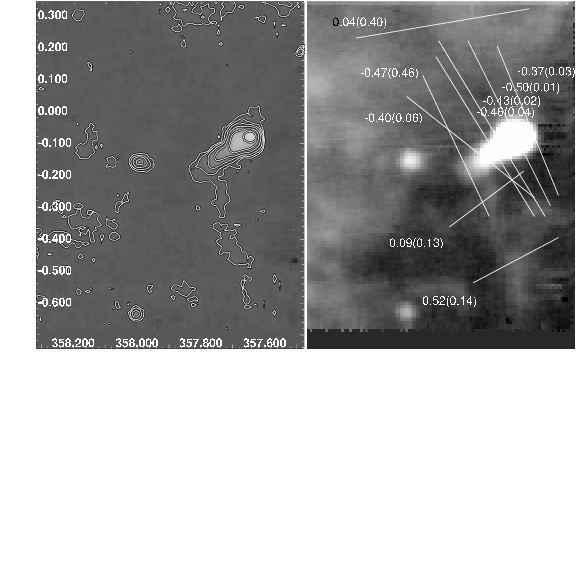}
\includegraphics[bb=0 200 600 600,clip,width=6.5in]{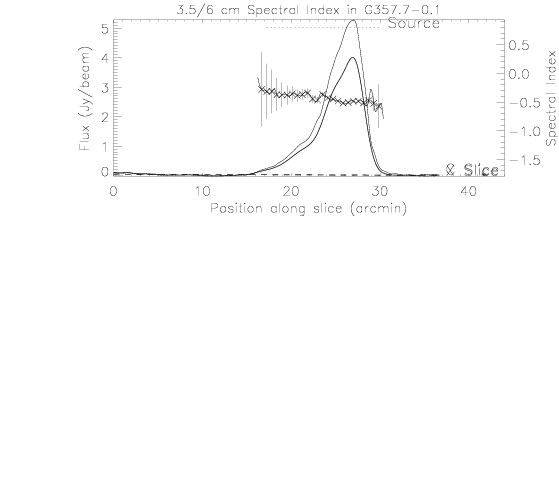}
\caption{\emph{Top, left}:  3.5 cm image of the region around G357.7--0.1 (Tornado) with contours at levels of $0.04*2^n$ Jy beam$^{-1}$, for $n=0-9$.  \emph{Top, right}:  Grayscale shows the 6 cm image of the same region with slices and corresponding spectral index at the peak brightness of each slice.  \emph{Bottom}:  Brightnesses and the corresponding spectral index are shown for the slice with $\alpha_{CX}=-0.48\pm0.02$ in the top right figure (along the long axis of the Tornado).  Note that comparing these spectral indices to other works should account for the flux calibration uncertainty by adding a spectral index error of 0.13 in quadrature.  \label{sptornado}}
\end{figure}

\begin{figure}[tbp]
\includegraphics[bb=0 200 600 600,clip,width=6.5in]{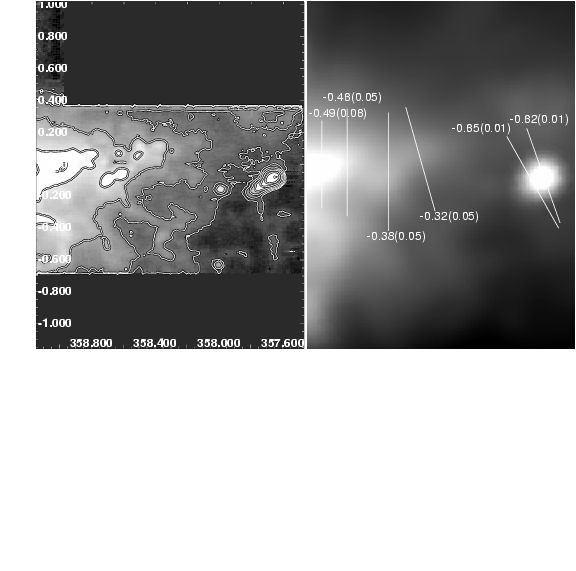}
\includegraphics[bb=0 200 600 600,clip,width=6.5in]{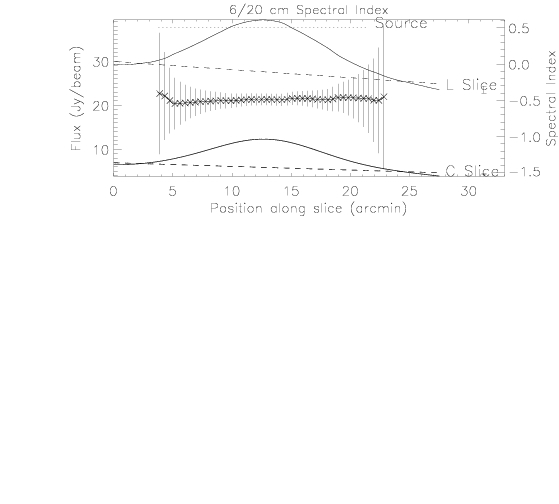}
\caption{Same as for Fig. \ref{sptornado}, but the left and right images show 6 and 20 cm emission in the west of the survey, respectively. Contours on the 6 cm survey are at $0.08*2^n$ Jy beam$^{-1}$, for $n=0-10$.  The plotted slice values correspond to the slice with $\alpha_{LC}=-0.49\pm0.06$ with the origin at the south.  \label{sptornadocl}}
\end{figure}

\begin{figure}[tbp]
\includegraphics[bb=0 200 600 600,clip,width=6.5in]{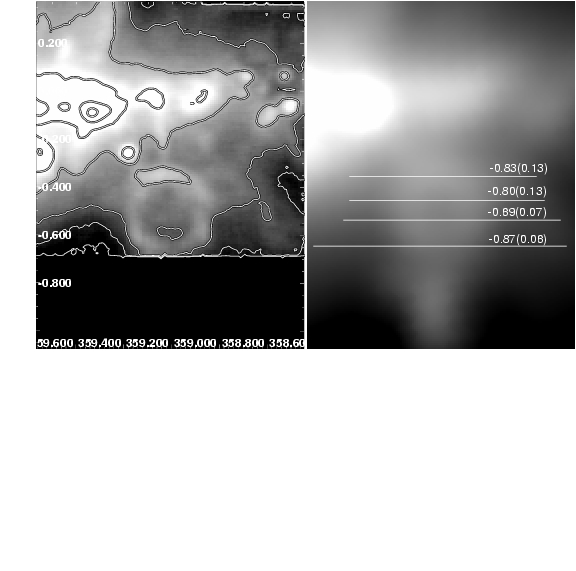}
\includegraphics[bb=0 200 600 600,clip,width=6.5in]{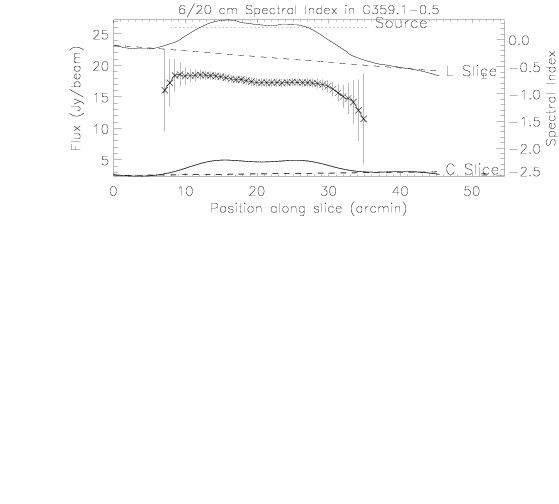}
\caption{Same as for Fig. \ref{sptornadocl}, showing images at 6 and 20 cm for the G359.1--0.5 SNR.  The plotted slice values correspond to the slice with $\alpha_{LC}=-0.89\pm0.07$ with origin at the east.  \label{spg359.1-0.5cl}}
\end{figure}

\begin{figure}[tbp]
\includegraphics[bb=0 200 600 600,clip,width=6.5in]{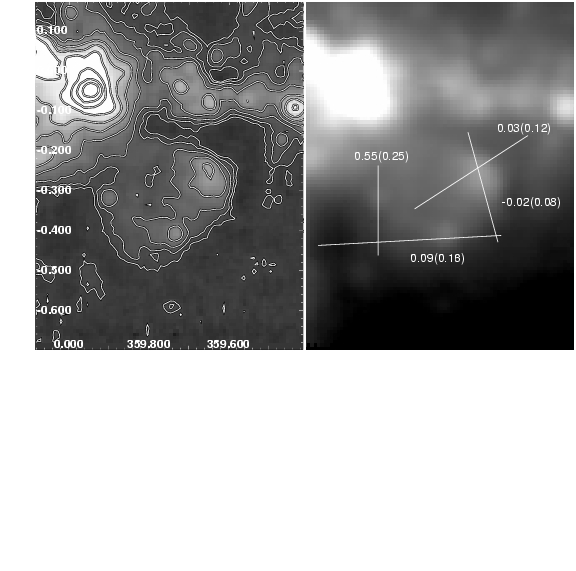}
\includegraphics[bb=0 200 600 600,clip,width=6.5in]{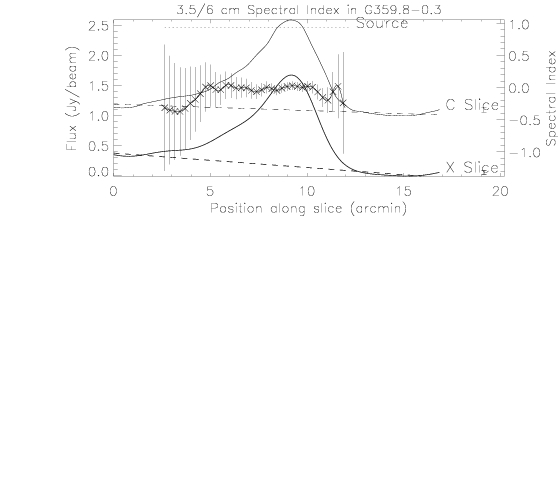}
\caption{Same as for Fig. \ref{sptornado}, but for the G359.8--0.3 complex.  The plotted slice values correspond to the slice with $\alpha_{CX}=0.03\pm0.12$ with origin at the southeast.  \label{spg359.8-0.3}}
\end{figure}

\begin{figure}[tbp]
\includegraphics[bb=0 200 600 600,clip,width=6.5in]{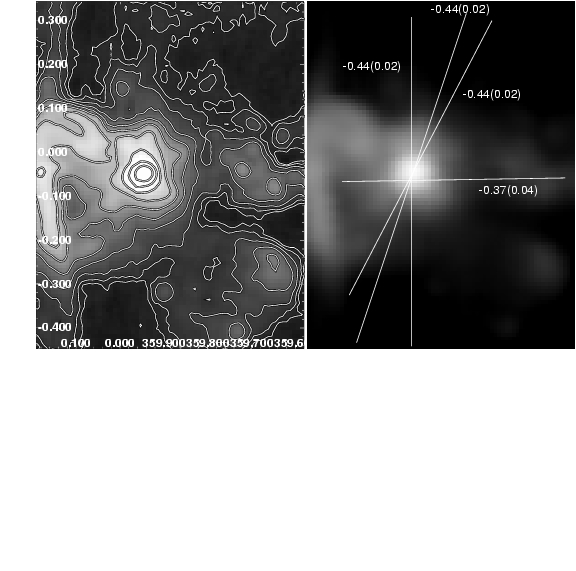}
\includegraphics[bb=0 200 600 600,clip,width=6.5in]{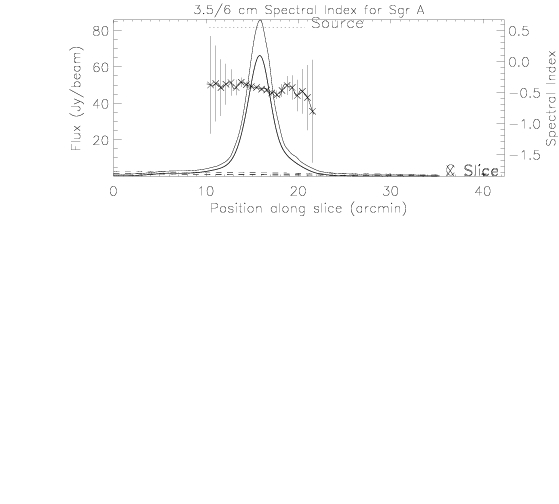}
\caption{Same as for Fig. \ref{sptornado}, but for the Sgr A complex (G0.07+0.04).  The plotted slice values correspond to the slice with $\alpha_{CX}=-0.44\pm0.02$ and the rightmost label with origin at the southeast.  \label{spsgra}}
\end{figure}

\begin{figure}[tbp]
\includegraphics[bb=0 200 600 600,clip,width=6.5in]{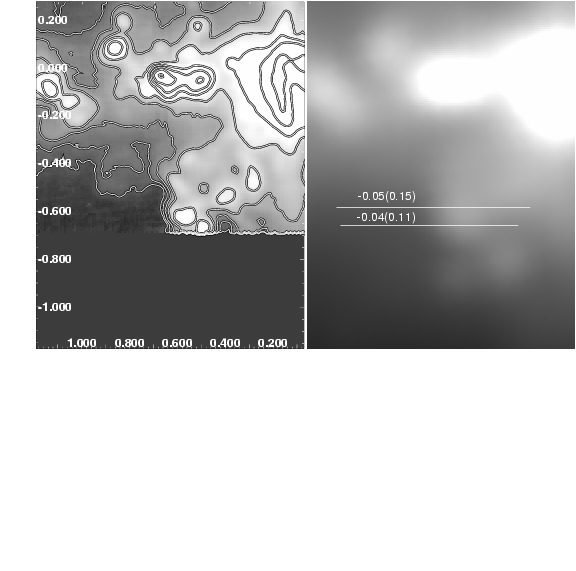}
\includegraphics[bb=0 200 600 600,clip,width=6.5in]{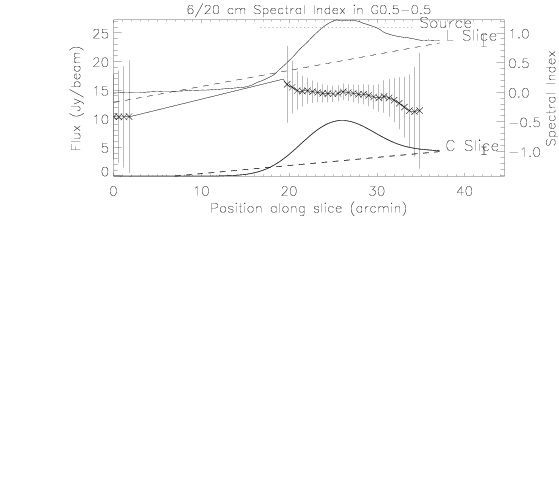}
\caption{Same as for Fig. \ref{sptornadocl}, showing images at 6 and 20 cm for the G0.5--0.5 complex. The plotted slice values correspond to the slice with $\alpha_{LC}=-0.04\pm0.11$ with origin at the east.  \label{spg0.5-0.5cl}}
\end{figure}

\begin{figure}[tbp]
\includegraphics[bb=0 200 600 600,clip,width=6.5in]{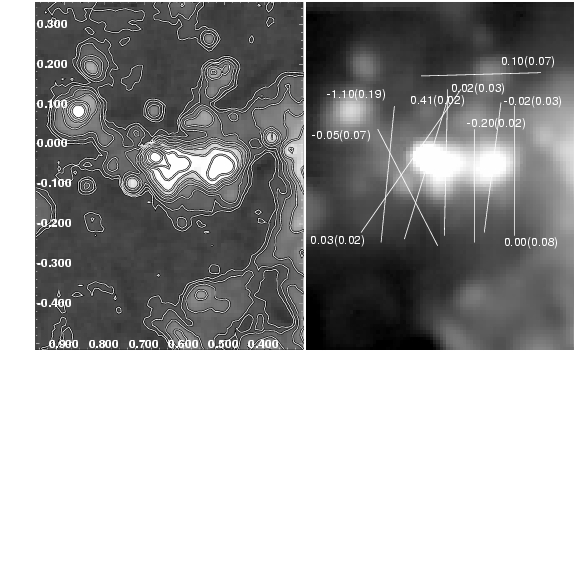}
\includegraphics[bb=0 200 600 600,clip,width=6.5in]{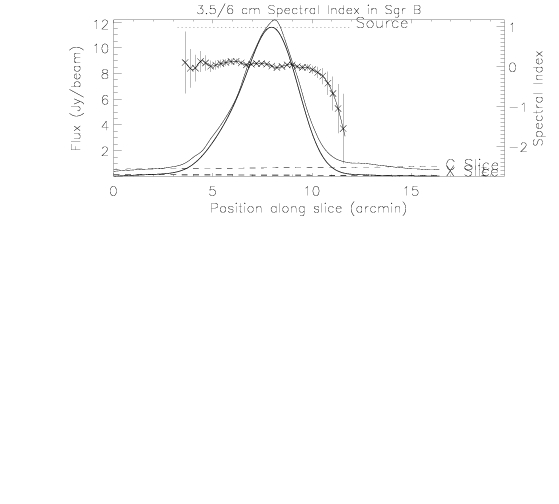}
\caption{Same as for Fig. \ref{sptornado}, but for the Sgr B complex (G0.5--0.0 and G0.7--0.0).  The plotted slice values correspond to the slice with $\alpha_{CX}=0.02\pm0.03$ with origin at the south.  \label{spsgrb}}
\end{figure}

\begin{figure}[tbp]
\includegraphics[bb=0 200 600 600,clip,width=6.5in]{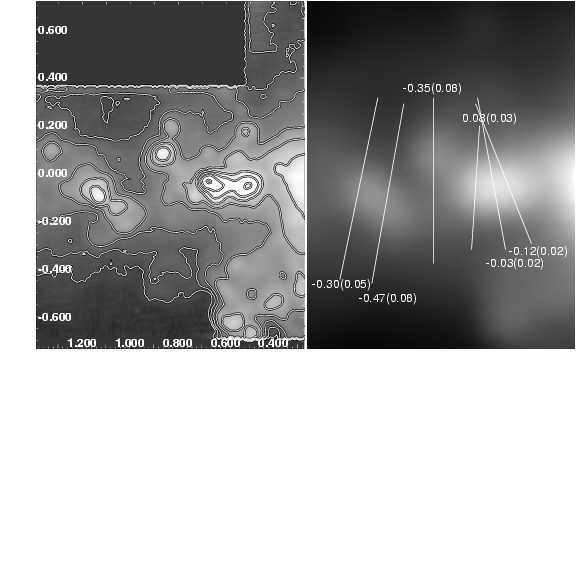}
\includegraphics[bb=0 200 600 600,clip,width=6.5in]{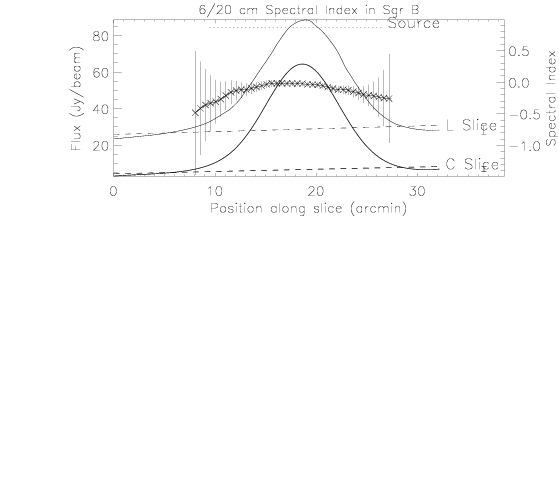}
\caption{Same as for Fig. \ref{sptornadocl}, showing images at 6 and 20 cm for the Sgr B (G0.5--0.0 and G0.7--0.0) complex and eastern sources. The plotted slice values correspond to the slice with $\alpha_{LC}=-0.03\pm0.02$ with origin at the north.  \label{spsgrbcl}}
\end{figure}

G357.7--0.1, also known as ``The Tornado'' for its unusual, twisted morphology, is found near the western edge of the 3.5 and 6 cm maps.  In Figure \ref{sptornado}, the 3.5 and 6 cm morphology looks like a head-tail source with integrated flux densities of about 14, 18, and 37 Jy at 3.5, 6 cm, and 20 cm, respectively.  The Tornado is believed to be a mixed-mophology SNR at a distance of about 12 kpc \citep{g03,b03a}.   Mixed-morphology SNRs are characterized by a radio continuum shell filled with thermal, x-ray--emitting gas \citep[e.g.,][]{yu03}.  The elongated morphology of the Tornado is unsual for a SNR, but can be explained by the fact that the Tornado is interacting with a molecular cloud \citep{f96}.

The spectral index between 6 and 3.5 cm is shown in Figure \ref{sptornado} and between 20 and 6 cm in Figure \ref{sptornadocl}.  The 6/3.5 cm spectral index for most slices perpendicular to the long axis are equal within their $3\sigma$ errors except for one slice on the western side (the ``head'') of the Tornado.  The typical spectral index values are $\alpha_{CX}\sim-0.45$ and $\alpha_{LC}\sim-0.63$, with systematic uncertainties of about 0.03 and 0.01, respectively.  The 6/3.5 cm spectral index from the slice analysis is consistent with the integrated spectral index of $-0.50\pm0.07$, given in Table \ref{diffsrcspix}.  Note that comparing these spectral indices to other works should account for the flux calibration uncertainty by adding a spectral index error of 0.13 in quadrature.  The 20/6 cm spectral index is steeper than the 6/3.5 cm index, which suggests that the spectral index steepens at lower frequencies.  From \citet{gr04} and \citet{b85}, the 1 GHz (30 cm) flux of the Tornado is 37 Jy and spectral index is --0.4, while \citet{g94} finds $S_{843 MHz}=49$ Jy.  Extrapolating from our measured 20 cm flux density of 37 Jy and $\alpha_{LC}\sim-0.63$, we predict an integrated flux density of 46 and 51 Jy at 1 GHz and 843 MHz (35 cm).  Thus, the present observations overpredict the flux given in \citet{gr04}, but are consistent with the observations of \citet{g94}.  The present single-dish--derived flux and spectral index are less likely to be biased than the previous interferometric values, which are more likely to underestimate the flux density.

The spectral index and flux density can be used to calculate an equipartition magnetic field.  \citet{b05} give a new derivation for the equipartition magnetic field strength as
\begin{equation}
B_{eq}=\left( \frac{4\pi (2\alpha+1) (K_0+1) I_\nu E_p^{1-2\alpha} (\nu/2c_1)^\alpha}{(2\alpha-1) c_2(\alpha) l c_4(i)}\right)^{1/(\alpha+3)}
\end{equation}
\noindent where $K_0$ is the proton to electron number density ratio, $c_i$ are constants that depend on the spectral index and magnetic field inclination angle, which hereafter is assumed to be equal to 0 (in the plane of the sky).  Alternatively, the classical formulation of the equipartition magnetic field is
\begin{equation}
B_{class}=(8\pi G (K+1) L_\nu/V)^{2/7}
\end{equation}
where $G$ is a function of the energy range considered and spectral index, $K$ is the energy density ratio between protons and electrons, and $V$ is the volume of the emitting region \citep{p70,b05}.

Assuming a proton to electron energy density ratio of 40-100 \citep{b05} and a path length through the Tornado equivalent to its 2\arcmin\ width (10.4 pc assuming $D=12$ kpc), the 6 cm peak brightness of the Tornado of 5 Jy beam$^{-1}$ and spectral index --0.48 gives $B_{class}=100-130\mu$G.  This calculation integrates over frequencies from 10 MHz (3 m) to 10 GHz (3 cm), but changes by less than 10\% for an upper limit of 100 GHz.

The spatial dependence of the spectral index is apparent in Figure \ref{sptornado}, which shows the flux density and spectral index for a slice through the elongated portion of the Tornado.  This slice shows a regular change in $\alpha_{CX}$ from $-0.48\pm0.02$ near the brightest emission to $-0.33\pm0.07$ in the tail.  The spectral index is a direct measure of the energy distribution of the electrons and suggests that that the electrons in the tail region are more energetic than in the head region.  

A new source, G357.7--0.4 is found near the Tornado in projection as shown in Figure \ref{sptornado}.  G357.7--0.4 is an elongated, wavy structure with a thermal spectral index.  The morphology and difference in spectral index suggests that it is unrelated to the Tornado.  The thermal emission from G357.7--0.4 should not affect the measurements of spectral index from the tail of G357.7--0.1, since they are significantly separated and oriented perpendicular to each other.

At the top of Figure \ref{sptornado} and north of the Tornado is G357.7+0.3, which has long been known as a supernova remnant from its ring-like morphology, linearly polarized radio emission, and soft X-ray emission \citep{r84,le89,g94}.  The slice across the southeast portion of G357.7+0.3 shown in Figure \ref{sptornado} has a flat spectral index with large uncertainty.  The spectral index is significantly steeper toward the southwest, since the 3.5 cm emission is absent, but the 6 cm brightness is similar to the southeast;  based on the 3.5 cm upper limit, we estimate $\alpha_{CX}\lesssim-1.5$ in the southwest of G357.7+0.3.  This work is consistent with previous work within their large uncertainties.  G357.7+0.3 is only half covered by the 3.5 and 6 cm surveys, and so the integrated flux density meausrements are not available.

\subsubsection{Sgr E Region (G358.7-0.0)}
\begin{figure}[tbp]
\includegraphics[bb=0 200 600 600,clip,width=6.5in]{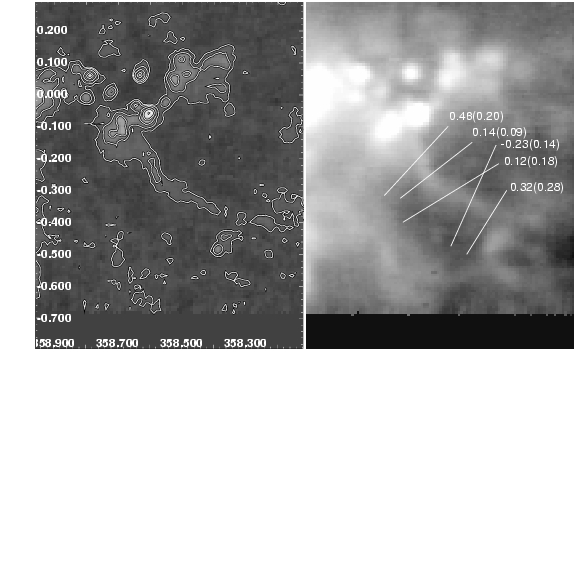}
\includegraphics[bb=0 200 600 600,clip,width=6.5in]{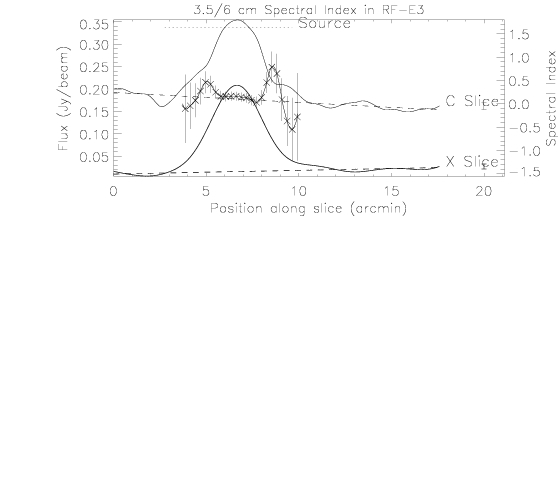}
\caption{Same as for Fig. \ref{sptornado}, but for the E3 radio filament (G358.60-0.27). The plotted slice values correspond to the slice with $\alpha_{CX}=0.12\pm0.18$ with the origin at the southeast. \label{spe3}}
\end{figure}

\begin{figure}[tbp]
\includegraphics[bb=0 200 600 600,clip,width=6.5in]{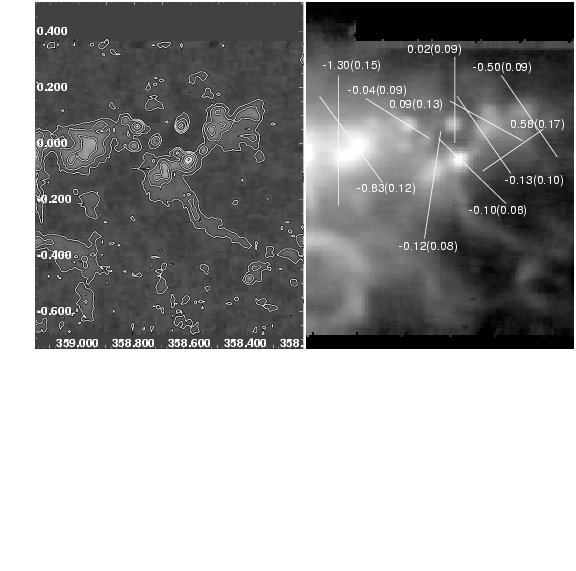}
\includegraphics[bb=0 200 600 600,clip,width=6.5in]{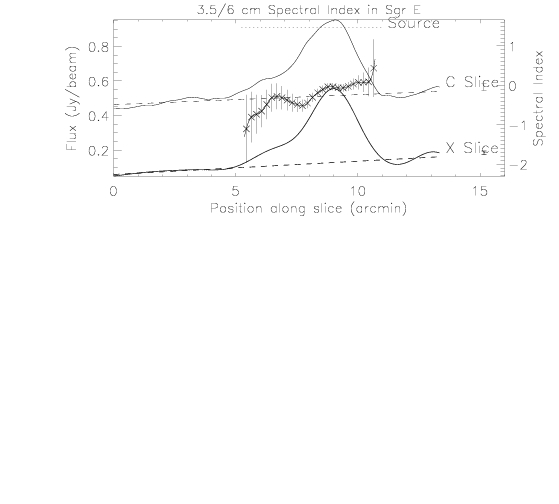}
\caption{Same as for Fig. \ref{sptornado}, but for the Sgr E complex (G358.7-0.0).  The plotted slice values correspond to the slice with $\alpha_{CX}=-0.04\pm0.09$ with the origin at the northeast. \label{spsgre}}
\end{figure}

Figures \ref{spe3} and \ref{spsgre} show the radio continuum emission from the E3 filament (G358.60-0.27) and the Sgr E complex (G358.7-0.0), respectively \citep{y04}.  The E3 filament is about 25\arcmin\ long and takes a twisting path from the Sgr E star-forming complex toward the southwest.  If the E3 filament is near the GC, it has a length of about 58 pc.  The Sgr E complex is made of a collection of compact sources within a 20\arcmin\ region near (358.7,0.0), surrounded by extended sources near (359.0,+0.0) and (358.4,+0.1).

The spectral index between 6 and 3.5 cm are shown in Figures \ref{spe3} and \ref{spsgre} and between 20 and 6 cm in Figure \ref{sptornadocl}.  High-resolution 20 cm continuum observations with the VLA show the region filled with compact \hii\ regions \citep{y04}, so much of the extended emission here could be unresolved compact sources.  Consistent with this expectation, the 6/3.5 cm spectral index for most of this emission is consistent with a thermal origin.  The E3 filament also has a thermal index throughout.

The easternmost and westernmost slices in Figure \ref{spsgre} show nonthermal 6/3.5 cm indices.  The easternmost slice passes through G359.0+0.0, which is seen in high-resolution, 20 cm images as extended, filamentary structures \citep{l92,y04}.  The morphology is remeniscent of an evolved \hii\ region, but the slice and integrated spectral indices are consistent with nonthermal emission.  The peak 6 cm brightness of 1 Jy beam$^{-1}$ and 6/3.5 cm spectral index of --0.63 implies a revised equipartition magnetic field strength of $65-85\mu$G, assuming it is in the GC and has an number density ratio of 40--100 \citep{b05}.

\subsubsection{SNR G359.1--0.5 and the Snake NRF (G359.1-0.2)}
\label{g359.1-0.5sec}
\begin{figure}[tbp]
\includegraphics[bb=0 200 600 600,clip,width=6.5in]{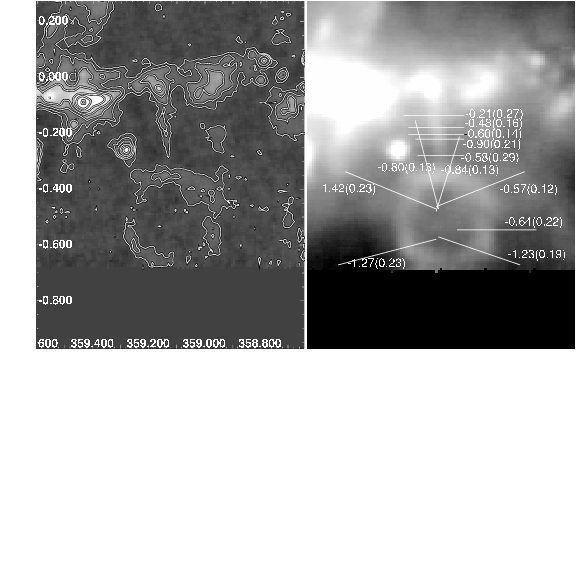}
\includegraphics[bb=0 200 600 600,clip,width=6.5in]{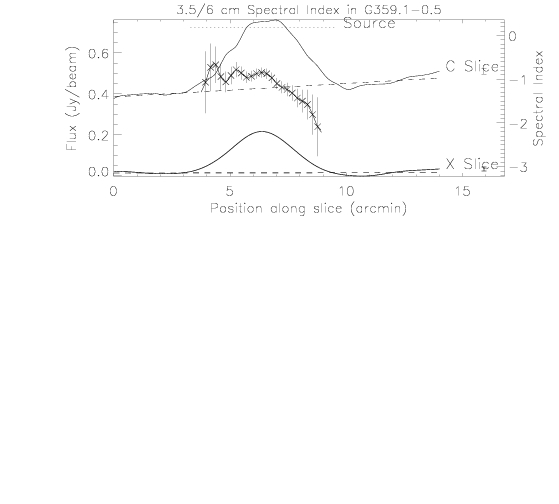}
\caption{Same as for Fig. \ref{sptornado}, but for G359.1--0.5 and the Snake (G359.1-0.2).  The plotted slice values correspond to the slice with $\alpha_{CX}=-0.84\pm0.13$ with origin at the south. \label{spg359.1-0.5}}
\end{figure}

Figure \ref{spg359.1-0.5} shows the 6 and 3.5 cm emission and slices across the SNR G359.1--0.5 and the G359.1--0.2 (also known as ``The Snake'').  G359.1--0.5 appears here as a ring of radius 9\damin5 (22 pc).  The ring of emission is noticably more irregular at 3.5 cm than at 6 cm, although it seems to be continuous at both wavelengths.  The flux density of G359.1--0.5 is $6.8\pm2.5$ Jy at 6 cm (see Table \ref{diffsrc}), which is consistent with the published value of $8.1\pm0.5$ \citep{r84}.  The SNR catalog of \citet{gr04} gives a 1 GHz flux density of 14 Jy and spectral index of approximately --0.4, which is also consistent with this flux density.  As shown in Table \ref{snrsrc}, the 6/3.5 cm spectral index in both our slice and integrated analysis methods range from --1.9 to --0.5, which is much steeper than that observed between 6 and 11 cm \citep{r84}.  The peak 6 cm brightness (0.3 Jy beam$^{-1}$; see Figure \ref{spg359.1-0.5}) and associated 6/3.5 cm spectral index (--0.84) are consistent with a revised equipartition magnetic field of $66-83\mu$G, assuming it is located near the GC and has a number density ratio of 40--100 \citep{b05}.

Studies of molecular gas and tracers of shocked gas have found that G359.1--0.5 is interacting with a molecular cloud \citep{u92}.  Interestingly, G359.1--0.5 has a statistically significant spatial variation in its spectral index.  Figure \ref{plsp} plots its 6/3.5 cm spectral index as a function of theta, the position on the ring of emission relative to galactic north.  The 20 cm image does not have the resolution to allow a similar study of the 20/6 cm spectral index.  The 6/3.5 cm spectral index for G359.1--0.5 is significantly flatter for $\theta=250-10$\sdeg\ (the galactic west through north side).  The region with flatter spectral index is nearly identical to the region with most intense emission from the surrounding HI, $^{12}$CO, and OH(1720 MHz) maser emission \citep{u92,y95}.  This is consistent with the idea that G359.1--0.5 is interacting with the surrounding molecular cloud \citep{u92}.

The Snake is a long ($\sim20$\arcmin\ or $\sim46$ pc at 8 kpc), nonthermal filament that runs from the Galactic plane to the G359.1--0.5 SNR \citep{u92,g95,y04}.  The Snake is unusual among NRFs because it has two sharp kinks along its length \citep[not visible with this data, but see][]{g95}.  The integrated flux densities given in Table \ref{diffsrc} imply a 6/3.5 cm spectral index of $-1.86\pm0.64$, which is steeper than the slice spectral index values.  Figure \ref{spg359.1-0.5} shows five slices across the Snake with spectral indices ranging from --0.2 to --0.9, tending to become more negative toward the south.  The slice analysis is more trustworthy than the integrated spectral index analysis, since this region is highly confused;  slice locations are chosen to avoid confusing sources.  The slice spectral index values are similar to that measured in interferometric observations between 20 and 6 cm \citep{g95}, although the previous work found some locations with a higher, even inverted, spectral index.  For a 6/3.5 cm spectral index of --0.6, a 6 cm brightness of 0.2 Jy beam$^{-1}$, and a depth equal to its width of 9\dasec4 \citep{g95}, the revised equipartition field strength is 137-176 $\mu$G, for $K_0=40-100$.  This field strength is somewhat larger than that of \citet{g95}, since we assume a larger value of $K_0$, which is more consistent with current estimates \citep{b05}.

\subsubsection{Sgr C Region (G359.5-0.0)}
\begin{figure}[tbp]
\includegraphics[bb=0 200 600 600,clip,width=6.5in]{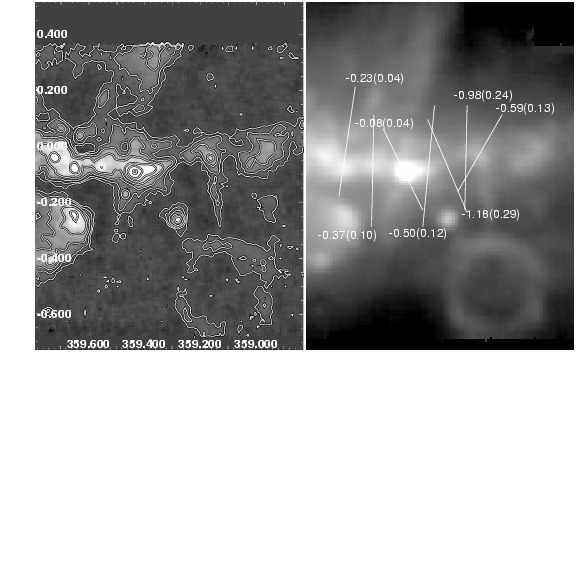}
\includegraphics[bb=0 200 600 600,clip,width=6.5in]{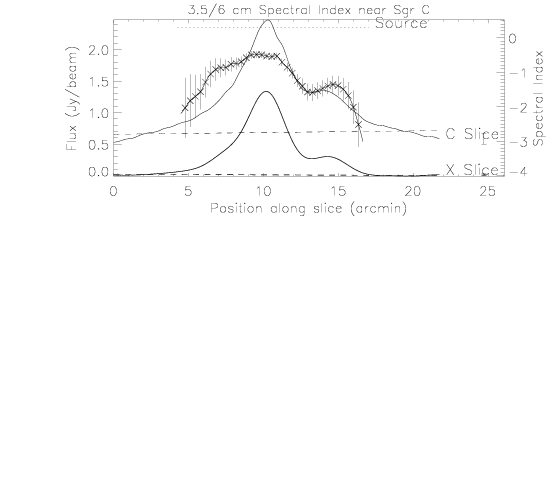}
\caption{Same as for Fig. \ref{sptornado}, but for the Sgr C complex (G359.5-0.0).  The plotted slice values correspond to the slice with $\alpha_{CX}=-0.59\pm0.13$ with origin at the south. \label{spsgrc}}
\end{figure}

Figure \ref{spsgrc} shows the 6 and 3.5 cm flux densities, slices, and spectral indices in the Sgr C complex.  The Sgr C \hii\ region around (359.5,--0.1) has flux densities of about 8 and 7 Jy at 3.5 and 6 cm, respectively, making it the brightest radio continuum source in the western half of the survey.  To the north of Sgr C near (359.4,+0.3) and possibly just south of Sgr C is the radio continuum counterpart of the GC lobe \citep{l08}.  G359.2+0.0 is located near where the Snake meets the Galactic plane.  This extended structure has flux densities of about 3 and 8 Jy at 3.5 and 6 cm, respectively.  

The slice and integrated flux densities show that nonthermal emission dominates the extended emission in the region, outside of the Sgr C \hii\ region.  G359.2+0.0 has a very steep spectral index between 6 and 3.5 cm.  The slice plotted in the bottom of Figure \ref{spsgrc} corresponds to the vertical slice just west of Sgr C;  it shows how the emission surrounding Sgr C is predominately nonthermal.

We note that some of the extended structure seen in the present survey is resolved into NRFs in high resolution observations \citep{y04,n04}.  The observed distribution of NRF flux density shows an increasing number of NRFs down to the present detection limits \citep{n04}.  More sensitive, high-resolution observations are likely to find many more faint NRFs than are currently observed.  Near Sgr C, the total 20 cm flux density of known NRFs and NRF candidates is roughly 2 Jy \citep{y04};  this is roughly 10\% of the total, background-subtracted, nonthermal 20 cm flux density observed by GBT of about 10--20 Jy.  We suggest that extended (but not Galactic synchrotron), nonthermal sources in the central degree of the GC region may have significant flux from NRFs.  While many other processes also produce nonthermal radio emission (e.g., SNRs), the example of the Sgr C region may help direct searches for NRFs;  other regions with similar radio continuum morphologies and spectral indices include G359.0+0.0, G359.2+0.0, and G0.8+0.0.

\subsubsection{G359.8--0.3}
Figure \ref{spg359.8-0.3} shows the continuum emission and slices in G359.8--0.3.  This source has a lumpy, shell-like morphology in this data, with flux densities of about 19 and 28 Jy at 3.5 and 6 cm, respectively.  The slices and integrated spectral indices of G359.8--0.3 are consistent with a thermal origin, suggesting that it is an \hii\ region.  Low-frequency radio continuum absorption and H$\alpha$ emission are observed from G359.8--0.3, as would be expected from a thermal source in the foreground to the GC region \citep{b03b,ga01}.  A \emph{Chandra} survey of the the region shows soft, X-ray emission associated with this shell, consistent with a foreground source \citep{w02}.

\subsubsection{Sgr A (SNR G0.0+0.0)}
Figures \ref{spsgra} and \ref{spsgracl} show the Sgr A region.  The peak radio brightness on arcminute scales in this region includes emission from Sgr A East and Sgr A West, the two brightest radio continuum sources in the central several arcminutes.  Sgr A East is roughly 4\arcmin$\times$3\arcmin, so no spatial structure is seen in the present survey and only the flux from the whole Sgr A complex is measured.  The peak brightness in this region is 39 Jy beam$^{-1}$ at 3.5 cm and 85 Jy beam$^{-1}$ at 6 cm.  Convolving the 3.5 cm map to the resolution of the 6 cm map gives a brightness of 66 Jy per 2\damin5 beam at 3.5 cm.  At 6 cm, VLA observations of the radio continuum find $70\pm10$ Jy comes from Sgr A East and $21\pm2$ Jy from Sgr A West, with a total flux density of about $88\pm10$ Jy \citep{b83,m89,p89}.  The results of the present survey are consistent with the interferometric observations, which suggests that they do not resolve much flux out \citep{p89}.

The 6/3.5 cm spectral index for the three slices across Sgr A, and for the integrated brightnesses given above are consistent with $\alpha_{CX}=-0.44\pm0.02$ for the central $\sim$2\damin5 (excluding systematic error).   The spectral index measured in the 6 and 20 cm slices give $\alpha_{LC}=-0.42\pm0.02$ for the central $\sim$9\arcmin.  High resolution observations of Sgr A East measure a spectral index of about --1.1 between 20 and 6 cm, which is believed to extend through the cm-wavelength regime \citep{p89}.  Sgr A West has a flat spectrum at wavelengths between 3.6 and 20 cm \citep{m89,p89}.  The combination of these two effects can explain the value of $\alpha_{CX}=-0.44$ observed in the present survey, although at longer wavelengths, the spectral index should become significantly steeper, since Sgr A East begins to dominate the apparent flux.  It is likely that the 6/20 cm spectral index measures a significant amount of the ``halo'' flux around the Sgr A complex, since the 20 cm map has a beam size of 9\arcmin.  The halo emission has a flat spectral index between 20 and 90 cm that may extend to shorter wavelengths \citep{p89}.

\subsubsection{Radio Arc (G0.2-0.0) and Arched Filaments (G0.07+0.04)}
\begin{figure}[tbp]
\includegraphics[bb=0 200 600 600,clip,width=6.5in]{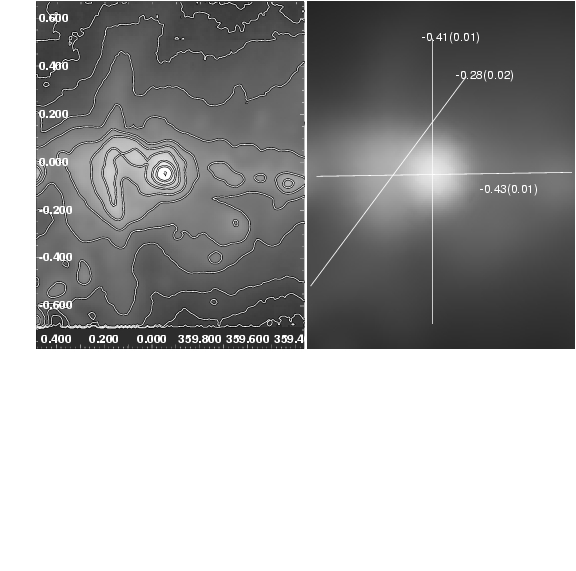}
\includegraphics[bb=0 200 600 600,clip,width=6.5in]{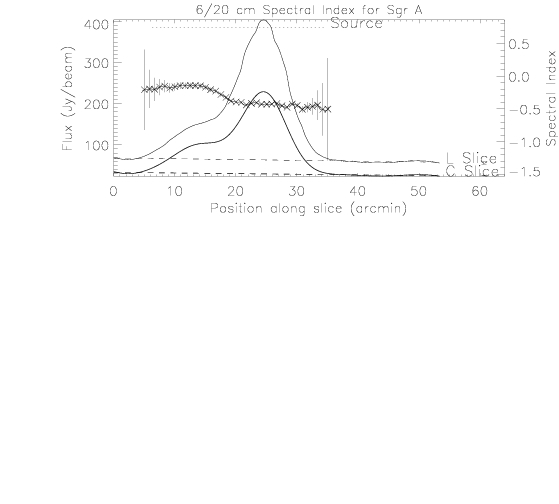}
\caption{Same as for Fig. \ref{sptornadocl}, showing images at 6 and 20 cm for the Sgr A complex (G0.0+0.0).  The plotted slice values correspond to the slice with $\alpha_{LC}=-0.43\pm0.01$ with origin at the east.  \label{spsgracl}}
\end{figure}

\begin{figure}[tbp]
\includegraphics[bb=0 200 600 600,clip,width=6.5in]{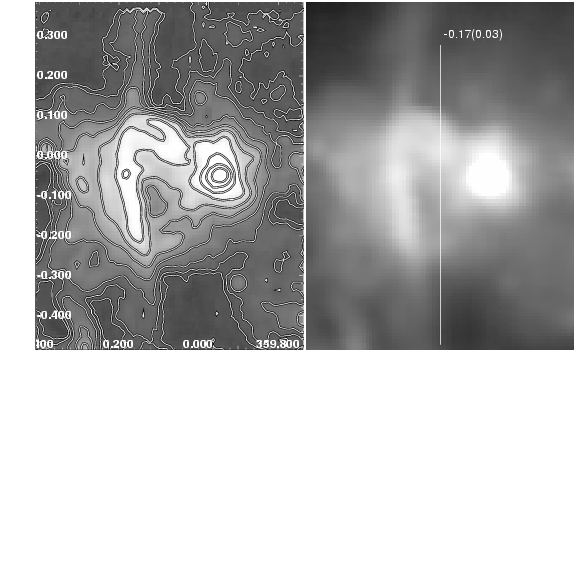}
\includegraphics[bb=0 200 600 600,clip,width=6.5in]{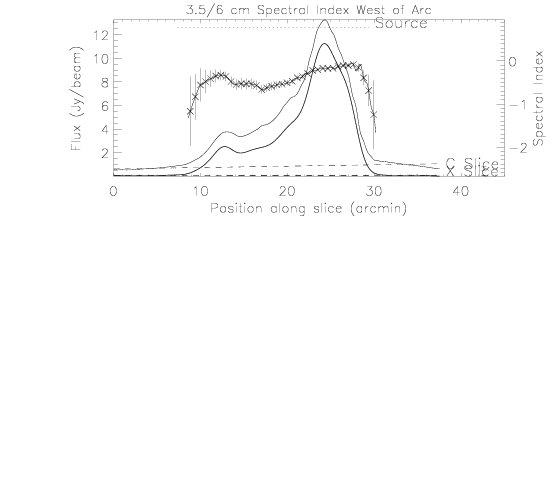}
\caption{Same as for Fig. \ref{sptornado}, but for the Arched filaments complex (G0.07+0.04).  The plotted slice values correspond to the slice shown at the top right with origin at the south. \label{spwestarc}}
\end{figure}

The Arched filaments/Radio Arc complex has been extensively studied at high resolution with the VLA \citep{y84,t86,la02,y02}.  The slices shown in Figures \ref{spsgracl}, \ref{spwestarc}, and \ref{speastarc} confirm the results of early observations that found that the Arched filaments are thermal and the vertical component of the Radio Arc is nonthermal.  The slice in Figure \ref{spwestarc} shows a slightly negative index, but this is likely due to a mixture of the thermal Arched filaments and the ambient nonthermal emission near the Radio Arc;  using a closer part of the slice to estimate the background of the Arched filaments gives $\alpha_{CX}\sim-0.1$.  The integrated arched filament fluxes and spectral index are difficult to evaluate due to confusion with surrounding emission, but are consistent with a thermal origin (see Table \ref{diffsrc}).  The Radio Arc spectral index is generally quite flat, as is shown in Figure \ref{spsgracl}, where $\alpha_{LC}=-0.14\pm0.04$.

\begin{figure}[tbp]
\plotone{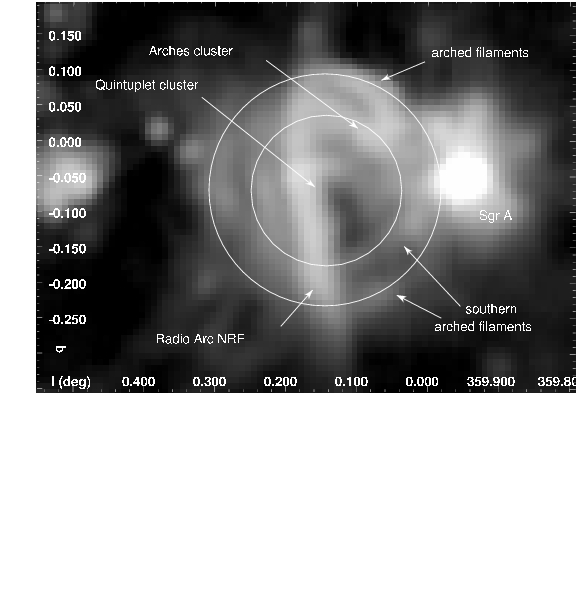}
\caption{GBT 3.5 cm image of the Radio Arc/arched filament region.  The positions of the Radio Arc (G0.2-0.0), Sgr A (G0.0+0.0), northern and southern arched filaments, and Arches and Quintuplet clusters are labeled.  The circles schematically show how the arched filaments (G0.07+0.04) connect to the southern arched filaments (G0.07--0.2) and other radio continuum structures to make a nearly continuous ring around the brightest portion of the Radio Arc;  this region is also referred to as the ``Radio Arc Bubble'' by \citet{si07}. \label{radioring}}
\end{figure}

The morphology of the radio continuum emission at 3.5 cm shows that the Arched filaments seem to be a part of a larger, ring-like structure sometimes referred to as ``the bubble'' \citep{r01,si07}.  Figure \ref{radioring} shows the GBT 3.5 cm image of the Arched filaments/Radio Arc region with the rings drawn schematically.  The Arched filaments can be connected to other radio continuum structures by two circles with similar center locations, near the brightest portion of the Radio Arc.  This ring-like structure has also been noted in filtered, 20 cm VLA images \citep{s03}.  The morphology of the Bubble in the present survey is similar to that of \citet{s03} and the unbiased, single-dish observations give some confidence in the identification of the ring structure.  The slice in Figure \ref{spwestarc} crosses the southwest portion of this ring and shows that the 6/3.5 cm spectral index is flatter there, suggesting there is a thermal contribution to the emission;  using a close background region shows that these ``southern Arched filaments'' (G0.07--0.2) have $\alpha_{CX}=0.11\pm0.13$, similar to the Arched filaments.

There is also a coincidence between the brightness of the Radio Arc and the extent of the Bubble.  As seen in Figure \ref{radioring}, the 3.5 cm emission from the Radio Arc is brightest inside the ring made by the Arched filaments and southern Arched filaments, but fades rapidly outside of that ring.  The brightness enhancement of the Radio Arc inside the Bubble is even more pronounced when compared to high Galactic latitutes (in the GC Lobe), where cm-wavelength continuum brightness is more than a factor of ten less than inside the Bubble \citep{si07,c07}.

\subsubsection{SNR G0.33+0.04}
\begin{figure}[tbp]
\includegraphics[bb=0 200 600 600,clip,width=6.5in]{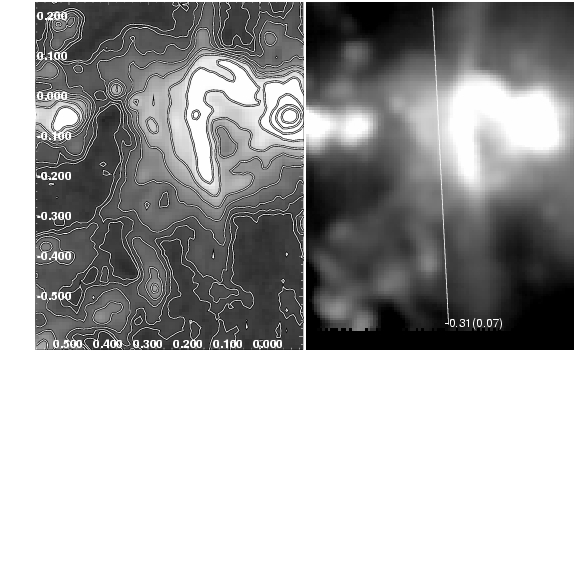}
\includegraphics[bb=0 200 600 600,clip,width=6.5in]{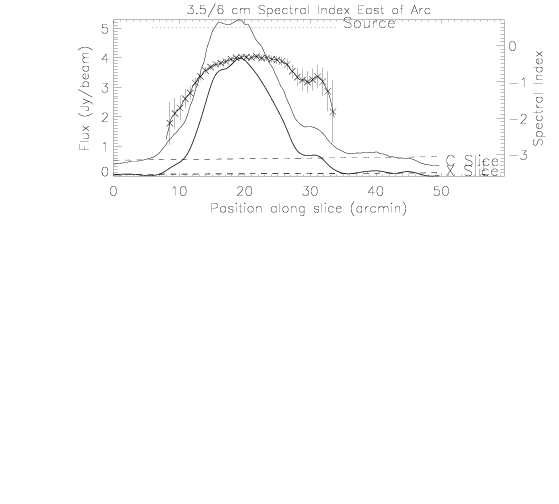}
\caption{Same as for Fig. \ref{sptornado}, but for the emission east of the Radio Arc (G0.2-0.0).  The plotted slice values correspond to the slice shown at the top right with origin at the north.  At position near 10\arcmin, the slice crosses the SNR G0.33+0.04 and has a spectral index $\alpha_{CX}$ of roughly --2.0 to --1.5.  \label{speastarc}}
\end{figure}

The SNR G0.33+0.04 is located adjacent to (and partially confused with) the Radio Arc in Figure \ref{speastarc}.  Because it is so confused with the Radio Arc emission, an integrated spectral analysis is not done.  A multiwavelength study of G0.33+0.04 finds a spectral index of --0.56 for frequencies higher than a GHz \citep{ka96}.  The slice shown in Figure \ref{speastarc} crosses a portion of this SNR and finds a spectral index $\alpha_{CX}=-2$ to $-1.5$, significantly steeper than previous measurements at 5 and 15.5 GHz (6 and 2 cm) \citep{a79,g74}.  However, we note that at these frequencies the emission is dominated by emission from the Radio Arc and a few \hii\ regions and the morphology does not resemble the clear SNR-like morphology seen at lower frequencies \citep{ka96}.  This confusion is likely to bias the flux densities measured by the present and previous works at frequencies above 5 GHz.

\subsubsection{G0.5--0.5}
\begin{figure}[tbp]
\includegraphics[bb=0 200 600 600,clip,width=6.5in]{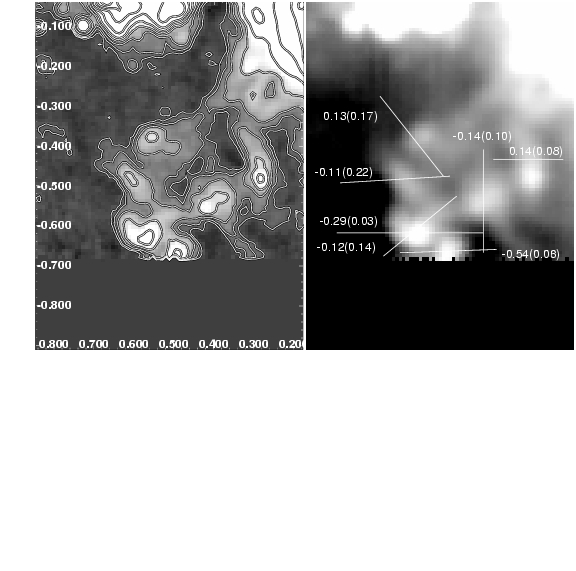}
\includegraphics[bb=0 200 600 600,clip,width=6.5in]{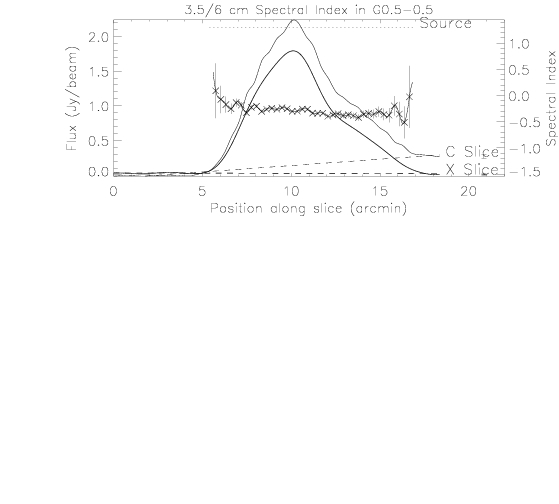}
\caption{Same as for Fig. \ref{sptornado}, but for the G0.5--0.5 complex.  The plotted slice values correspond to the slice with $\alpha_{CX}=-0.29\pm0.03$ with origin at the east.  \label{spg0.5-0.5}}
\end{figure}

Figure \ref{spg0.5-0.5} shows G0.5--0.5, an irregular complex of clumpy extended emission with peak brightnesses around 2 Jy beam$^{-1}$ at 6 cm.  The 6/3.5 cm spectral index measurements shown in Figure \ref{spg0.5-0.5} are near zero, showing that G0.5--0.5 is predominately thermal.  Like G359.8--0.3, this region is seen in absorption in 74 MHz continuum and emission in optical H$\alpha$, indicating that it is in the foreground of the GC region \citep{b03b,ga01}.  One slice in Figure \ref{spg0.5-0.5} shows a significantly nonthermal spectrum, although it appears morphologically similar to the rest of the G0.5--0.5 complex.  The integrated and slice spectral index values are consistent with nonthermal emission in the southern portion of G0.5--0.5.  All other 20/6 cm spectral index measurements across G0.5--0.5 shown in Figure \ref{spg0.5-0.5cl} are consistent with a thermal origin.

\subsubsection{Sgr B (G0.5--0.0 and G0.7--0.0)}
As shown in Figures \ref{spsgrb} and \ref{spsgrbcl}, Sgr B is the brightest radio continuum source in the eastern half of the survey.  The emission is brightest in two regions, called Sgr B1  (G0.5--0.0) and Sgr B2 (G0.7--0.0), the western and eastern halves, respectively.  The entire Sgr B region has flux densities of 90 Jy and 85 Jy at 3.5 and 6 cm, respectively, of which roughly 60\% comes from Sgr B2.  Sgr B is also surrounded by a few isolated \hii\ regions that are detected as compact sources in Tables \ref{srcX} and \ref{srcC}.

The 6/3.5 cm and 20/6 cm spectral index measurements for most of the Sgr B complex are near zero, which is consistent with it being thermal.  As shown in Table \ref{diffsrc}, the 6 and 3.5 cm spectral index values are near zero with three exceptions:  one significantly larger than zero and two much less than zero.  The positive spectral index is measured at the brightest portion of Sgr B2, a region known to be filled with optically-thick, ultracompact \hii\ regions \citep{ga95}.  The nonthermal spectral index measured east of Sgr B2 is in a region with extended emission that is not clearly associated with the Sgr B complex.  The other nonthermal spectral index is measured between Sgr B1 and B2.  Sgr B1 has more extended radio continuum morphology and has fewer compact \hii\ regions that Sgr B2, which suggests that the region is relatively older than Sgr B2 \citep{m92}.  The nonthermal index measured between the two, at (0.55,--0.05), may indicate that a supernova has occurred there, although no clear morphological signatures of a supernova in the Sgr B complex have ever been reported.  \citet{k07} recently reported an extended region with 6.7 keV thermal line emission at (0.61,--0.01), which they interpret as a young ($7\times10^3$ yr-old) supernova remnant.  The 6.7 keV-emitting region and the nonthermal radio emission are both located between Sgr B1 and B2 and only about 4\arcmin\ from each other, so the radio continuum emission is consistent with the supernova hypothesis.  The nonthermal radio index measured between Sgr B1 and Sgr B2 is --0.2, flatter than that of SNRs, but it is likely to be mixed with \hii\ regions.  The observation of nonthermal continuum emission in the Sgr B complex is consistent with previous observations from throughout the electromagnetic spectrum \citep{y07,h07,cr07}.  Observations of 90 cm continuum find a total Sgr B flux density of $\sim31$Jy \citep{l00}, giving a spectral index of $\sim0.35$, perhaps indicative of significant absorption by thermal gas in the region.

\subsubsection{Sgr D (G1.1--0.1), SNR G1.0--0.2, and SNR G0.9+0.1}
\begin{figure}[tbp]
\includegraphics[bb=0 200 600 600,clip,width=6.5in]{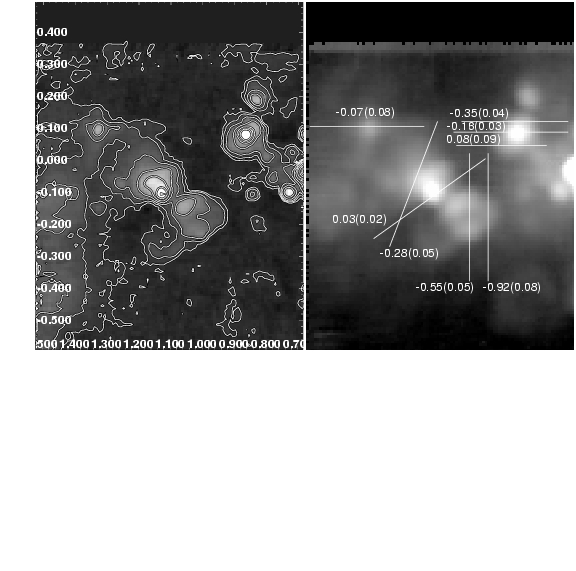}
\includegraphics[bb=0 200 600 600,clip,width=6.5in]{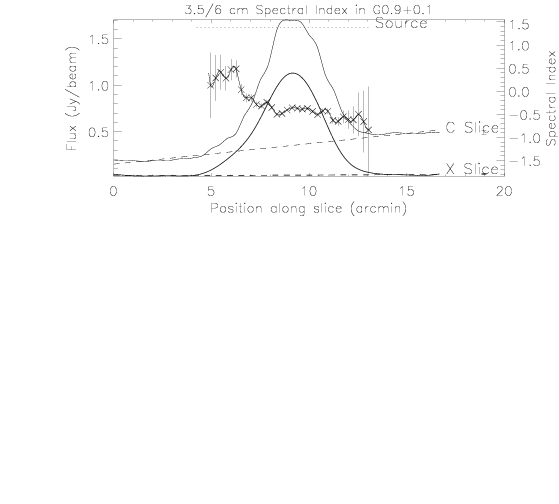}
\caption{Same as for Fig. \ref{sptornado}, but for G0.9+0.1 and eastern complex.  The plotted slice values correspond to the slice with $\alpha_{CX}=-0.35\pm0.04$ with origin at the east.  \label{spg0.9+0.1}}
\end{figure}

The eastern edge of the survey has a mixture of thermal and nonthermal sources, as seen in the images and slices in Figures \ref{spsgrbcl} and \ref{spg0.9+0.1}.  Sgr D (called ``G1.1--0.1'' in Table \ref{snrsrc}) appears here as a compact source (at 3.5 cm; see Table \ref{srcX}) surrounded by a shell.  Previous work has suggested that the Sgr D shell is ionized by the compact source at its center and that the extended radio continuum emission toward the east (called ``G1.2+0.0'' here) may be gas escaping from Sgr D \citep{l92}.  G0.9+0.1 and G1.0--0.2 \citep[a.k.a. ``G1.05--0.1''][]{g94} are supernova remnants with clear shell-like structures and nonthermal 6/3.5 and 20/6 cm spectral indices.  G0.9+0.1 also has a Crab-like source inside the SNR (with a similar flux density as the shell at 6 cm), making it one of the first SNRs categorized as a ``composite'' \citep{h87}.

The integrated spectral index of the Sgr D extended emission of $\alpha_{CX}=-0.22\pm0.14$ is consistent with its identification as an ionized shell of gas.  Although the slice through G1.1--0.1 (the extended Sgr D emission) is nominally nonthermal ($\alpha_{CX}=-0.28\pm0.05$), parts of that slice have $\alpha_{CX}=-0.18\pm0.10$, which suggests that there is a mixture of thermal and nonthermal emission in the region.  The slice through the compact source at (1.1,--0.1) clearly has a thermal-like index of $0.03\pm0.02$, which is consistent with previous work that identified that source as an \hii\ region with radio recombination line emission with $v_{LSR}\approx25$ \kms \citep[source G1.127-1.04 of][]{l92}.  The source is detected as a compact source at 3.5 cm at (17:48:40.29,--28:01:23.5) with a peak brightness of 2.3 Jy and flux density of 3.5 Jy, as compared to the flux density near 19 cm of 1.7 Jy \citep{l92}.

G0.9+0.1 is known to have a flat spectrum pulsar wind nebula in its center and a steep-spectrum shell \citep{h87,g94}.  The slices shown in Figure \ref{spg0.9+0.1} are generally consistent with this picture, showing flat 6/3.5 cm indices in the core ($\sim-0.16$) and a steeper spectral index in the shell ($\sim-0.35$).  Although the compact source inside the SNR has a flat spectral index, it also has linear polarization, which is consistent with suggestions that it is a pulsar wind nebula \citep{h87}.  The peak brightness for the unresolved core emission is about 4.6 and 4.0 Jy per 2\damin5-beam at 6 and 3.5 cm, respectively, which is similar to the 6 cm flux density of 4.16 Jy observed by \citet{h87}; the present observation is also consistent with the 90 cm flux density of $4.8\pm0.07$ Jy \citep{l00} considering the flat spectral index of the core.  The flux density of the shell is difficult to estimate, since it is confused with the core in the 2\damin5-resolution images;  the shell brightness is about 1.3 and 1.1 Jy per 2\damin5-beam, at 6 and 3.5 cm, respectively.  The flux density of the shell is given in Table \ref{diffsrc} as 9.08 Jy and 5.46 Jy, at 6 and 3.5 cm, respectively.  The 6/3.5 cm spectral indices for the core and shell from the slice analysis are $-0.16\pm0.03$ and $-0.35\pm0.04$, respectively, which is consistent with the values of $-0.1$ and $-0.45$ found by \citet{h87}.  In the shell, comparing the 6 cm flux density to the 90 cm flux density of \citet{l00} implies a slightly flatter index of $\sim0.2$, possibly indicating a spectral turnover.

G1.0--0.2 is a supernova remnant with a shell morphology and nonthermal radio spectral index \citep[called G1.05--0.1 in][]{g94}.  Unlike G0.9+0.1, there is no compact source inside G1.0--0.2.  Figure \ref{spg0.9+0.1} shows two slices through G1.0--0.2 with $\alpha_{CX}=-0.55\pm0.05$ and $-0.92\pm0.08$, consistent with previous observations that found an index that ranged from $-0.6$ to $-0.7$ between 57.5 MHz and 1616 MHz \citep{l92}.  Table \ref{diffsrc} shows the measured flux densities of G1.0--0.2 are 8.52 and 9.90 Jy at 3.5 and 6 cm, respectively, and an integrated spectral index $\alpha_{CX}=-0.27\pm0.18$, which is roughly similar to the slice results.  \citet{l92} report that G1.0--0.2 has a flux density of 12.3 Jy from interferometric observations at 1616 MHz; this gives an upper limit to the spectral index of $\alpha_{LC}<-0.2$.

\section{Conclusions}
\label{gcsurvey_con}
This paper has shown results from a new survey of the radio continuum emission from the central degrees of the Galaxy at 90, 20, 6, and 3.5 cm with the GBT.  The 6 and 3.5 cm surveys are the most sensitive, highest resolution, single-dish radio surveys of the central degrees of the GC made at these wavelengths.  The primary products of this study are catalogs of all compact and extended sources, including a spectral index analysis of these sources.  We have shown that the compact sources ($\theta\lesssim5$\arcmin) detected at 6 and 3.5 cm surveys are most likely to be Galactic \hii\ regions and are mostly found near well-known \hii\ complexes, such as Sgr B and Sgr E.  About one quarter of the sources detected at 3.5 cm are also detected at 6 cm;  most of these sources have thermal spectral indices.

Extended, nonthermal emission in the 3.5, 6, and 20 cm surveys is found associated with GC star-forming regions.  The emission is on size scales of tens of arcminutes and is particularly found for $l=358\ddeg5-359\ddeg5$.  The brightness-distribution function of NRFs estimated by high resolution observations indicates that they are much more numerous at low flux densities, so the extended nonthermal emission may yet be resolved as new, nonthermal radio filaments.  Nonthermal emission also traces supernova remnants, and these observations find nonthermal emission in Sgr B, consistent with a recent report of a supernova in that star-forming region.  Another interesting result from our spectral index study is that the 6/3.5 cm spectral index distribution around the G359.1--0.5 SNR is consistent with the idea that it is interacting with a neighboring molecular cloud that may be in the GC region.

We also find a structure south of the Arched filaments that is thermal and seems to be morphologically connected to the well-known Arched filaments.  The Arched filaments and the new southern Arched filaments (G0.07--0.2) form a contiguous structure that is a radio-continuum counterpart to the Radio Arc Bubble.  The Bubble surrounds the brightest part of the nonthermal emission from the Radio Arc and is close to the dense star clusters, the Quintuplet and Arches clusters.  Several recent works has definitively shown that the ionization of the Arched filaments is caused by the Arches star cluster, while the rest of the Bubble is ionized by the Quintuplet star cluster \citep{r01,la02,si07}.  \citet{si07} further suggests that the Bubble was formed by the winds and ionizing photons of massive stars the Quintuplet cluster.  This work highlights the fact that the Bubble --- a thermal structure formed by stellar action --- has a clear effect on the brightness of the nonthermal Radio Arc.  The reason for the enhanced brightness inside the Bubble is not clear, but should be considered in models for the formation of NRFs.

Through a combination of slice and integrated flux analysis, all objects detected in the surveys are catagorized as having a thermal or nonthermal origin.  Thus, the distribution of flux between thermal and nonthermal sources can be quantified.  Within the central 4\sdeg$\times$1\sdeg\ of the Galaxy, the thermal to nonthermal flux fractions for all discrete emission are 28\%/72\% at 3.5 cm and 19\%/81\% at 6 cm.  This does not include the background synchrotron contribution from the Galactic plane, which begins to dominate the Galaxy's flux density for wavelengths longer than 6 cm.  Also, some of these sources in the field are likely to be in the foreground of the GC region, although the density of gas and stars is generally much higher in the GC region, so most sources are likely to truly be in the central few hundred parsecs of the Galaxy.  The high fraction of nonthermal emission in the radio continuum emission is consistent with the idea that the cosmic ray density is enhanced in the nuclear disk, assuming that the magnetic field is not unusually strong ($\sim1$ mG) in the region.  Other studies have also found evidence for such a cosmic ray enhancement, which seems heat and ionize molecular gas in the GC region, possibly affecting the star formation process \citep{y07}.

\begin{acknowledgements}
Thanks to James Miller-Jones for help with estimating extragalactic source counts.  The National Radio Astronomy Observatory is a facility of the National Science Foundation operated under cooperative agreement by Associated Universities, Inc.
\end{acknowledgements}

{\it Facilities:} \facility{GBT ()}

\clearpage

\begin{deluxetable}{ccccccc}
\tablecaption{Overview of GBT Surveys of GC Region \label{mapstats}}
\tablewidth{0pt}
\tablehead{
\colhead{Band} & \colhead{$\lambda$} & \colhead{$\nu$} & \colhead{Resolution} & \colhead{Long. Range} & \colhead{Lat. Range} & \colhead{Sensitivity\tablenotemark{b}} \\
\colhead{} & \colhead{(cm)} & \colhead{(GHz)} & \colhead{(arcmin)} & \colhead{(deg)} & \colhead{(deg)} & \colhead{(mJy beam$^{-1}$)} \\
}
\startdata
X & 3.5 & 8.50 & 1.5 & 357.5, +1.5 & --0.7, +0.35 & 9 \\
C & 6.2 & 4.85 & 2.5 & 357.5, +1.5\tablenotemark{a} & --0.7, +0.35\tablenotemark{a} & 20 \\
L & 21.3 & 1.42 & 9.0 & 355.5, +7.6 & --5.1, +3.4 & 300 \\
P & 92.3 & 0.325 & 38.8 & 356.4, +7.6 & --5.4, +5.4 & 4000 \\
\enddata
\tablenotetext{a}{Coverage also extends up to $b=+0\ddeg8$ for $l=359\ddeg0$ to 0\sdeg5}
\tablenotetext{b}{Measured away from obvious sources, but tends to include background synchrotron emission at longer wavelengths.}

\end{deluxetable}

\begin{deluxetable}{ccccccccccccc}
\tablecaption{3.5 cm Compact Source Catalog for GC Region \label{srcX}}
\tabletypesize{\scriptsize}
\tablewidth{0pt}
\tablehead{
\colhead{\#} & \colhead{l} & \colhead{b} & \colhead{RA} & \colhead{Dec} & \colhead{$S_p$\tablenotemark{a}} & \colhead{$\sigma_{S_p}$\tablenotemark{a}} & \colhead{$S_i$} & \colhead{$\sigma_{S_i}$} & \colhead{bmaj} & \colhead{bmin} & \colhead{bpa} & \colhead{$S_i^{H92}$} \\ 
\colhead{} & \colhead{(deg)} & \colhead{(deg)} & \colhead{(J2000)} & \colhead{(J2000)} & \colhead{(Jy/bm)} & \colhead{(Jy/bm)} & \colhead{(Jy)} & \colhead{(Jy)} & \colhead{(arcsec)} & \colhead{(arcsec)} & \colhead{(deg)} & \colhead{(Jy)} \\ 
}
\startdata
1 & 1.128    &  --0.100 & 17:48:40.29 & --28:01:23.5 & 2.29  & 1.7E-2 & 3.48  & 4.0E-2 & 115.2 & 102.3 &  24 & $8.61\pm$20\% \\
2 & 0.864    &   +0.081 & 17:47:20.97 & --28:09:20.7 & 2.22  & 1.7E-2 & 4.35  & 4.7E-2 & 127.7 & 119.0 & 115 & $7.15\pm$10\% \\
3 & 0.843    &  --0.105 & 17:48:01.39 & --28:16:13.0 & 0.20  & 1.8E-2 & 0.21  & 3.3E-2 &  94.2 &  89.8 & 106 & \\
4 & 0.728    &  --0.099 & 17:47:43.64 & --28:21:55.3 & 1.29  & 1.7E-2 & 1.72  & 3.7E-2 & 104.0 &  98.4 &  67 & \\
5 & 0.719    &   +0.022 & 17:47:14.24 & --28:18:36.1 & 0.15  & 1.8E-2 & 0.18  & 3.5E-2 & 101.2 &  92.6 &  40 & \\
6 & 0.675    &   +0.083 & 17:46:53.71 & --28:18:59.5 & 0.31  & 1.7E-2 & 0.52  & 4.3E-2 & 119.0 & 110.0 &   5 & \\
7 & 0.671    &  --0.034 & 17:47:20.50 & --28:22:51.6 &17.59  & 1.7E-2 &28.10  & 4.1E-2 & 128.5 &  96.2 &  67 & $34.9\pm$10\% \\
8 & 0.605    &  --0.195 & 17:47:48.66 & --28:31:14.0 & 0.09  & 1.7E-2 & 0.17  & 4.6E-2 & 134.4 & 108.9 &  35 & \\
9 & 0.538    &   +0.264 & 17:45:52.33 & --28:20:21.3 & 0.29  & 1.7E-2 & 0.38  & 3.6E-2 & 104.9 &  96.4 &  44 & \\
10& 0.530    &   +0.133 & 17:46:21.55 & --28:24:51.5 & 0.19  & 1.7E-2 & 0.27  & 3.8E-2 & 122.7 &  89.5 & 134 & \\
11& 0.524    &   +0.178 & 17:46:10.24 & --28:23:46.2 & 0.66  & 1.7E-2 & 1.07  & 4.1E-2 & 116.2 & 108.6 &  96 & $1.94\pm$20\% \\
12& 0.475    &   +0.066 & 17:46:29.41 & --28:29:47.6 & 0.13  & 1.7E-2 & 0.16  & 3.6E-2 & 122.9 &  80.2 &  74 & \\
13& 0.380    &   +0.017 & 17:46:27.43 & --28:36:09.1 & 1.20  & 1.8E-2 & 1.27  & 3.2E-2 &  92.8 &  88.7 &  15 & \\
14& 0.326    &  --0.015 & 17:46:27.26 & --28:39:57.1 & 0.90  & 1.8E-2 & 0.95  & 3.2E-2 &  93.3 &  87.8 & 121 & \\
15& 359.896  &  --0.319 & 17:46:37.11 & --29:11:27.2 & 0.22  & 1.7E-2 & 0.42  & 4.5E-2 & 125.0 & 115.6 &  30 & \\
16& 359.782  &   +0.035 & 17:44:57.84 & --29:06:12.4 & 0.14  & 1.7E-2 & 0.39  & 5.9E-2 & 216.8 &  97.3 &  53 & \\
17& 359.742  &  --0.594 & 17:47:19.94 & --29:27:55.3 & 0.07  & 1.7E-2 & 0.12  & 4.3E-2 & 139.9 &  94.9 &  80 & \\
18& 359.722  &  --0.040 & 17:45:06.72 & --29:11:40.0 & 0.61  & 1.7E-2 & 1.06  & 4.3E-2 & 137.5 &  97.9 &  49 & \\
19& 359.694  &   +0.003 & 17:44:52.63 & --29:11:42.9 & 0.38  & 1.7E-2 & 0.77  & 4.8E-2 & 138.4 & 113.7 &  32 & \\
20& 359.628  &   +0.039 & 17:44:34.60 & --29:13:59.3 & 0.22  & 1.6E-2 & 0.77  & 7.2E-2 & 193.3 & 142.1 & 134 & \\
21& 359.467  &  --0.173 & 17:45:01.28 & --29:28:51.6 & 0.22  & 1.8E-2 & 0.23  & 3.1E-2 &  97.5 &  81.8 & 172 & \\
22& 359.433  &   +0.001 & 17:44:15.64 & --29:25:10.0 & 0.16  & 1.6E-2 & 0.62  & 7.6E-2 & 218.6 & 134.6 & 179 & \\
23& 359.281  &  --0.261 & 17:44:55.23 & --29:41:09.7 & 0.89  & 1.7E-2 & 1.71  & 4.6E-2 & 125.4 & 119.1 & 167 & $1.84\pm$10\% \\
24& 358.801  &  --0.009 & 17:42:46.24 & --29:57:45.2 & 0.14  & 1.8E-2 & 0.17  & 3.5E-2 & 106.3 &  90.8 &  71 & \\
25& 358.789  &   +0.059 & 17:42:28.52 & --29:56:13.3 & 0.40  & 1.7E-2 & 0.52  & 3.6E-2 & 108.4 &  91.7 &  83 & $0.6\pm$30\% \\
26& 358.723  &   +0.008 & 17:42:30.98 & --30:01:10.2 & 0.15  & 1.8E-2 & 0.17  & 3.4E-2 & 107.5 &  82.6 & 148 & \\
27& 358.690  &  --0.087 & 17:42:48.30 & --30:05:52.3 & 0.19  & 1.7E-2 & 0.52  & 5.9E-2 & 154.7 & 137.1 &  10 & \\
28& 358.689  &  --0.125 & 17:42:57.26 & --30:07:08.1 & 0.19  & 1.7E-2 & 0.32  & 4.3E-2 & 127.2 & 103.4 & 176 & \\
29& 358.646  &  --0.036 & 17:42:29.95 & --30:06:31.3 & 0.15  & 1.8E-2 & 0.17  & 3.3E-2 & 106.5 &  80.8 &  53 & \\
30& 358.631  &   +0.062 & 17:42:04.58 & --30:04:10.5 & 0.41  & 1.8E-2 & 0.49  & 3.4E-2 & 104.6 &  89.0 & 157 & $0.45\pm$20\% \\
31& 358.605  &  --0.062 & 17:42:30.11 & --30:09:25.7 & 0.57  & 1.7E-2 & 0.98  & 4.3E-2 & 126.4 & 104.4 & 142 & $1.18\pm$10\% \\
32& 358.548  &  --0.022 & 17:42:12.44 & --30:11:04.3 & 0.10  & 1.8E-2 & 0.13  & 3.5E-2 & 130.5 &  73.7 & 140 & \\
33& 358.383  &  --0.481 & 17:43:36.98 & --30:33:57.3 & 0.12  & 1.7E-2 & 0.25  & 4.9E-2 & 154.3 & 105.9 & 129 & \\
34& 358.187  &   +0.302 & 17:40:02.83 & --30:19:09.3 & 0.05  & 1.7E-2 & 0.13  & 5.5E-2 & 159.0 & 121.2 &  91 & \\
35& 358.003  &  --0.635 & 17:43:17.72 & --30:58:12.3 & 0.24  & 1.8E-2 & 0.29  & 3.4E-2 &  97.6 &  93.5 &   0 & \\
36& 357.992  &  --0.160 & 17:41:23.32 & --30:43:44.0 & 0.40  & 1.7E-2 & 0.70  & 4.3E-2 & 134.8 &  99.7 &  81 & \\
\enddata
\tablenotetext{a}{GBT beam FWHM at 3.5 cm is 88\arcsec.}

\end{deluxetable}

\begin{deluxetable}{cccccccccccccc}
\tablecaption{6 cm Compact Source Catalog for GC Region \label{srcC}}
\tabletypesize{\scriptsize}
\tablewidth{0pt}
\tablehead{
\colhead{\#} & \colhead{l} & \colhead{b} & \colhead{RA} & \colhead{Dec} & \colhead{$S_p$\tablenotemark{a}} & \colhead{$\sigma_{S_p}$\tablenotemark{a}} & \colhead{$S_i$} & \colhead{$\sigma_{S_i}$} & \colhead{bmaj} & \colhead{bmin} & \colhead{bpa} \\ 
\colhead{} & \colhead{(deg)} & \colhead{(deg)} & \colhead{(J2000)} & \colhead{(J2000)} & \colhead{(Jy/bm)} & \colhead{(Jy/bm)} & \colhead{(Jy)} & \colhead{(Jy)} & \colhead{(arcsec)} & \colhead{(arcsec)} & \colhead{(deg)} \\ 
}
\startdata
1 & 0.862 &  +0.081   & 17:47:20.61 & --28:09:25.4 & 3.81 & 1.5E-1 & 7.35 & 4.2E-1 & 224.0 & 200.8 & 123 \\
2 & 0.724 & --0.090   & 17:47:41.07 & --28:21:51.9 & 1.79 & 1.5E-1 & 3.81 & 4.5E-1 & 223.4 & 212.8 & 138 \\
3 & 0.670 & --0.034   & 17:47:20.33 & --28:22:51.2 & 21.73& 1.6E-1 & 32.8 & 3.6E-1 & 199.5 & 176.5 & 68  \\
4 & 0.673 &  +0.084   & 17:46:53.31 & --28:19:05.4 & 0.54 & 1.5E-1 & 1.49 & 5.5E-1 & 266.4 & 243.0 & 113  \\
5 & 359.282 & --0.258 & 17:44:54.68 & --29:40:59.6 & 1.34 & 1.6E-1 & 1.79 & 3.3E-1 & 177.9 & 175.9 & 92  \\
6 & 358.791 & +0.063  & 17:42:27.76 & --29:55:59.2 & 0.45 & 1.6E-1 & 0.52 & 3.1E-1 & 192.2 & 139.0 & 78  \\
7 & 358.630 & +0.066  & 17:42:03.57 & --30:04:05.8 & 0.46 & 1.6E-1 & 0.57 & 3.2E-1 & 182.0 & 157.5 & 84  \\
8 & 358.606 & --0.061 & 17:42:30.04 & --30:09:20.1 & 0.86 & 1.5E-1 & 1.60 & 4.1E-1 & 239.6 & 180.9 & 116 \\
9 & 358.004 & --0.634 & 17:43:17.56 & --30:58:07.7 & 0.29 & 1.6E-1 & 0.29 & 2.8E-1 & 160.5 & 148.1 & 51  \\
10& 357.990 & --0.158 & 17:41:22.48 & --30:43:47.2 & 0.54 & 1.6E-1 & 0.77 & 3.5E-1 & 197.7 & 168.3 & 94  \\
\enddata
\tablenotetext{a}{GBT beam FWHM at 6 cm is 153\arcsec.}

\end{deluxetable}

\begin{deluxetable}{ccccc}
\tablecaption{Spectral Index for Compact Sources Detected at 3.5 and 6 cm \label{psspix}}
\tablewidth{0pt}
\tablehead{
\colhead{\#} & \colhead{l} & \colhead{b} & \colhead{$\alpha_{CX}$} & \colhead{$\sigma_\alpha$\tablenotemark{a}} \\ 
\colhead{} & \colhead{(deg)} & \colhead{(deg)} & \colhead{} & \colhead{} \\ 
}
\startdata
1 & 0.862 &  +0.081   & --0.93 & 0.10 \\
2 & 0.724 & --0.090   & --1.42 & 0.21 \\
3 & 0.670 & --0.034   & --0.28 & 0.02 \\
4 & 0.673 &  +0.084   & --1.88 & 0.67 \\
5 & 359.282 & --0.258 & --0.08 & 0.33 \\
6 & 358.791 & +0.063  &   0.00 & 1.07 \\
7 & 358.630 & +0.066  & --0.27 & 1.01 \\
8 & 358.606 & --0.061 & --0.87 & 0.46 \\
9 & 358.004 & --0.634 &   0.00 & 1.73 \\
10& 357.990 & --0.158 & --0.17 & 0.82 \\
\enddata
\tablenotetext{a}{Error in spectral index based on statistical errors and do not account for absolute flux calibration errors.  For the expected 5\% absolute flux errors in the 3.5 and 6 cm maps, a spectral index uncertainty of $\sim0.13$ should be added in quadrature to these errors.}

\end{deluxetable}

\begin{deluxetable}{cccccc|cccc|cccc}
\tablecaption{Extended Source Catalog for 3.5 and 6 cm GBT Observataions of GC Region \label{diffsrc}}
\tabletypesize{\scriptsize}
\tablewidth{0pt}
\tablehead{
 & & & & & \multicolumn{4}{c}{3.5 cm} & \multicolumn{4}{c}{6 cm} \\
\hline
\colhead{Name} & \colhead{l} & \colhead{b} & \colhead{RA} & \colhead{Dec} & \colhead{$R_{eff}$} & \colhead{raw $S_{i,X}$} & \colhead{$\sigma_{bg,X}$} & \colhead{$S_{i,X}$} & \colhead{$\sigma_{S_{i,X}}$} & \colhead{raw $S_{i,C}$} & \colhead{$\sigma_{bg,C}$} & \colhead{$S_{i,C}$} & \colhead{$\sigma_{S_{i,C}}$} \\ 
 & \colhead{(deg)} & \colhead{(deg)} & \colhead{(J2000)} & \colhead{(J2000)} & \colhead{(arcmin)} & \colhead{(Jy)} & \colhead{(Jy/bm)} & \colhead{(Jy)} & \colhead{(Jy)} & \colhead{(Jy)} & \colhead{(Jy/bm)} & \colhead{(Jy)} & \colhead{(Jy)} \\ 
}
\startdata
Tornado & 357.66 & --0.09 & 17:40:17.0 & --30:58:12 &      6.5 & 14.16 & 9.7E-3 & 13.81 & 0.51 & 18.45 & 1.6E-2     & 18.27 & 0.41 \\ 
G357.7--0.4 & 357.70 & --0.44 &  17:41:46.6 & -31:07:35 & 7.0 & 2.09 & 1.0E-2 & 1.55 & 0.71   & 1.30 & 1.6E-2      & 1.05 & 0.55 \\ 
G358.4+0.1 & 358.38 & +0.12 & 17:41:15.9 & --30:15:01 &    4.0 & 1.52 & 7.8E-3 & 1.24 & 0.14   & 3.41    & 1.9E-2   & 1.85 & 0.16 \\
RF E3 & 358.54 & --0.29 & 17:43:14.1 & --30:19:56 &        6.5 & 1.99 & 9.0E-3 & 1.51 & 0.45   & 4.28 & 2.3E-2      & 1.99 & 0.56 \\ 
Sgr E th\tablenotemark{a} & 358.50 & +0.05 & 17:41:49.0 & --30:11:09 &      8.0 & 4.54 & 8.0E-3 & 3.67 & 0.61   & 12.12 & 7.2E-2     & 3.79 & 2.64 \\ 
G359.0+0.0 & 358.91 & --0.03 & 17:43:07.9 & --29:52:54 &   8.0 & 8.13 & 9.3E-3 & 7.15 & 0.71   & 31.06   & 5.1E-2   & 13.52 & 1.86 \\
G359.1--0.5 & 359.11 & --0.51 & 17:45:29.3 & --29:57:35 & 12.0 & 6.24 & 1.0E-2 & 3.29 & 1.84   & 28.84 & 4.0E-2     & 6.81 & 2.46 \\ 
Snake & 359.14 & --0.19 & 17:44:17.9 & --29:46:00 &        5.0 & 1.21 & 1.1E-2 & 1.05 & 0.35   & 7.97 & 3.6E-2      & 2.99 & 0.55 \\ 
G359.2+0.0 & 359.18 & +0.01 & 17:43:37.5 & --29:38:04 &    5.5 & 3.24 & 1.0E-2 & 2.81 & 0.37   & 13.47 & 3.6E-2     & 7.71 & 0.64 \\ 
GCL-NW & 359.38 & +0.27 & 17:43:04.1 & --29:19:26 &        8.0 & 6.10 & 1.0E-2 & 5.07 & 0.85   & 19.38 & 5.2E-2     & 7.59 & 2.13 \\ 
Sgr C th\tablenotemark{a} & 359.44 & --0.10 & 17:44:39.6 & --29:27:54 &   2.5 & 7.56 & 8.2E-3 & 7.51 & 0.05   & 8.77 & 1.2E-1      & 6.82 & 0.34 \\ 
GCL-SW & 359.49 & --0.28 & 17:45:31.3 & --29:31:05 &       4.5 & 0.89 & 1.0E-2 & 0.51 & 0.27   & 7.40 & 1.8E-1      & 2.00 & 2.30 \\ 
Sgr C nt\tablenotemark{b} & 359.56 & --0.07 & 17:44:50.9 & --29:21:16 &     9.5 & 22.78 & 8.2E-3 & 21.79 & 0.99 & 67.69 & 1.2E-1     & 28.19 & 6.96 \\ 
G359.8--0.3 & 359.77 & --0.32 & 17:46:19.3 & --29:18:04 & 10.5 & 20.31 & 1.5E-2 & 18.721 & 2.12 & 56.08 & 1.8E-1     & 27.57 & 12.20 \\ 
Sgr A & 359.95 & --0.06 & 17:45:42.9 & --29:00:39 & 17.5 & 332.18\tablenotemark{c} & 1.3E-2 & 289.14 & 5.54 & 529.07\tablenotemark{c} & 9.1E-2 & 461.17 & 18.57 \\ 
Arched filaments & 0.09 & +0.05 & 17:45:39.6 & --28:49:53    &       6.0 & 66.80 & 6.2E-1 & 37.53 & 24.55 & 84.23 & 1.3E+0     & 8.81 & 24.66 \\ 
Arc & 0.18 & --0.06 & 17:46:16.9 & --28:48:52       &     16.5 & 175.34 & 1.4E-2 & 171.31 & 4.59& 277.25 & 9.1E-2    & 231.87 & 14.28 \\
G0.5--0.5 th\tablenotemark{a} & 0.42 & --0.49 & 17:48:31.7 & --28:49:33 &  12.5 & 22.08 & 1.2E-2 & 20.39 & 2.39 & 51.73 & 1.2E-1     & 22.82 & 11.47 \\ 
G0.5--0.5 nt\tablenotemark{b} & 0.57 & --0.63 & 17:49:24.0 & --28:46:32 &   5.5 & 8.01 & 1.2E-2 & 7.72 & 0.40   & 14.78 & 4.7E-2     & 14.60 & 0.76 \\ 
Sgr B & 0.62 & --0.06 & 17:47:18.9 & --28:25:58 & 8.5 & 103.60\tablenotemark{d} & 1.1E-2 & 72.23 & 1.07 & 123.34\tablenotemark{d} & 1.2E-1 & 64.67 & 5.59 \\ 
G0.8+0.0 & 0.81 & --0.03 & 17:47:38.9 & --28:15:42 &       4.0 & 2.58 & 9.1E-3 & 2.42 & 0.20   &  9.51 & 2.8E-2     & 6.42 & 0.29 \\
G0.8+0.2 & 0.82 & +0.20 & 17:46:46.1 & --28:07:44 &        3.5 & 2.90 & 1.5E-2 & 2.44 & 0.25   &  4.91 & 1.1E-1     & 2.99 & 1.48 \\
G0.8--0.4 & 0.83 & --0.42 & 17:49:14.1 & --28:26:24 &      7.5 & 1.49 & 1.2E-2 & 0.95 & 0.84   &  3.00 & 2.7E-2     &  1.98 & 0.90 \\ 
G0.9+0.1 & 0.87 & +0.08 & 17:47:22.5 & --28:09:19 &        4.5 & 5.91 & 1.3E-2 & 5.46 & 0.35   & 12.46 & 4.1E-2     & 9.08 & 1.61 \\
G1.0--0.2 & 1.02 & --0.17 & 17:48:42.5 & --28:09:11 &      5.5 & 9.25 & 1.5E-2 & 8.52 & 0.53   & 15.95 & 7.0E-2     &  9.90 & 1.18 \\
G1.1--0.3 & 1.13 & --0.28 & 17:49:22.5 & --28:06:55 &      5.0 & 2.50 & 1.5E-2 & 1.92 & 0.41   &  4.99 & 2.9E-2     &  2.37 & 0.38 \\
G1.1--0.1 & 1.13 & --0.07 & 17:48:33.4 & --28:00:06 &      3.5 & 5.04 & 1.1E-2 &  4.93 & 0.15  &  6.84 & 9.2E-2     &  5.59 & 0.58 \\
G1.2+0.0 & 1.23 & +0.01 & 17:48:28.9 & -27:52:37 &        8.5 & 10.99 & 1.1E-2 & 10.17 & 1.09 & 18.71 & 9.2E-2     & 9.33 & 4.37 \\
\enddata
\tablenotetext{a}{The ``th'' appended to the source name emphasizes that the source region is defined for the thermal-emitting part of the complex.}
\tablenotetext{b}{The ``nt'' appended to the source name emphasizes that the source region is defined for the nonthermal-emitting part of the complex.}
\tablenotetext{c}{Flux includes Arched filaments extended source, which is subtracted with the background.}
\tablenotetext{d}{Flux includes the Sgr B2 point source, which is subtracted with the background.}

\end{deluxetable}

\begin{deluxetable}{cccccc}
\tablecaption{Spectral Indices for Extended Sources \label{diffsrcspix}}
\tabletypesize{\scriptsize}
\tablewidth{0pt}
\tablehead{
\colhead{Name} & \colhead{l} & \colhead{b} & \colhead{$\alpha_{CX}$} & \colhead{$\sigma_{\alpha}$\tablenotemark{a}} & \colhead{Thermal/Nonthermal} \\ 
 & \colhead{(deg)} & \colhead{(deg)} & & & \\ 
}
\startdata
Tornado     & 357.66 & --0.09 & --0.50 & 0.07 & NT \\ 
G357.7--0.4 & 357.70 & --0.44 &  +0.69 & 1.04 & T \\ 
G358.4+0.1  & 358.38 &  +0.12 & --0.71 & 0.23 & NT \\
Sgr E th\tablenotemark{b}    & 358.50 &  +0.05 & --0.06 & 0.91 & T \\ 
RF E3       & 358.54 & --0.29 & --0.49 & 0.63 & T \\ 
G359.0+0.0  & 358.91 & --0.03 & --1.23 & 0.25 & NT \\
G359.1--0.5 & 359.11 & --0.51 & --1.30 & 1.19 & NT \\ 
Snake       & 359.14 & --0.19 & --1.86 & 0.64 & NT \\ 
G359.2+0.0  & 359.18 &  +0.01 & --1.80 & 0.26 & NT \\ 
GCL-NW      & 359.38 &  +0.27 & --0.72 & 0.46 & NT \\ 
Sgr C th\tablenotemark{b}    & 359.44 & --0.10 &  +0.17 & 0.06 & T \\ 
GCL-SW      & 359.49 & --0.28 & --2.44 & 1.70 & NT \\ 
Sgr C nt\tablenotemark{c}    & 359.56 & --0.07 & --0.46 & 0.32 & NT \\ 
G359.8--0.3 & 359.77 & --0.32 & --0.69 & 0.58 & T \\ 
Sgr A       & 359.95 & --0.06 & --0.83 & 0.06 & NT \\ 
Arched filaments &   0.09 &  +0.05 &  +2.58 & 3.65 & T \\ 
Arc         &   0.18 & --0.06 & --0.54 & 0.09 & NT \\
G0.5--0.5 th\tablenotemark{b} &  0.42 & --0.49 & --0.20 & 0.66 & T \\ 
G0.5--0.5 nt\tablenotemark{c} &  0.57 & --0.63 & --1.14 & 0.11 & NT \\ 
Sgr B       &   0.62 & --0.06 &  +0.20 & 0.11 & T \\ 
G0.8+0.0    &   0.81 & --0.03 & --1.74 & 0.16 & NT \\
G0.8+0.2    &   0.82 &  +0.20 & --0.36 & 0.64 & NT \\
G0.8--0.4   &   0.83 & --0.42 & --1.31 & 1.67 & T \\ 
G0.9+0.1    &   0.87 &  +0.08 & --0.91 & 0.25 & NT \\
G1.0--0.2   &   1.02 & --0.17 & --0.27 & 0.18 & NT \\
G1.1--0.3   &   1.13 & --0.28 & --0.38 & 0.43 & NT \\
G1.1--0.1   &   1.13 & --0.07 & --0.22 & 0.14 & T \\
G1.2+0.0    &   1.23 &  +0.01 &  +0.15 & 0.60 & T \\
\enddata
\tablenotetext{a}{Error in spectral index based on statistical errors and do not account for absolute flux calibration errors.  For the expected 5\% absolute flux errors in the 3.5 and 6 cm maps, a spectral index uncertainty of $\sim0.13$ should be added in quadrature to these errors.}
\tablenotetext{b}{The ``th'' appended to the source name emphasizes that the source region is defined for the thermal-emitting part of the complex.}
\tablenotetext{c}{The ``nt'' appended to the source name emphasizes that the source region is defined for the nonthermal-emitting part of the complex.}

\end{deluxetable}

\begin{deluxetable}{ccccccccccc}
\tablecaption{Catalog of Known and Candidate Supernova Remnants Observed at 3.5, 6, and 20 cm \label{snrsrc}}
\tabletypesize{\scriptsize}
\tablewidth{0pt}
\tablehead{
\colhead{Name} & \colhead{RA} & \colhead{Dec} & \colhead{Size} & \colhead{$S_{X}$} & \colhead{$S_{C}$} & \colhead{$\alpha_{CX}^{\rm{int}}$} & \colhead{$\alpha_{CX}^{\rm{slice}}$} & \colhead{$\alpha_{LC}^{\rm{slice}}$} & \colhead{Figures\tablenotemark{a}} & \colhead{References\tablenotemark{b}} \\ 
               & \colhead{(J2000)} & \colhead{(J2000)} & \colhead{(arcmin)} & \colhead{(Jy)} & \colhead{(Jy)} & & & & & \\ 
}
\startdata
G357.7--0.1 (Tornado)\tablenotemark{c} & 17:41:46.6 & -31:07:35 & 4$\times$2.5 & $13.81\pm0.51$ & $18.27\pm0.28$ & $-0.50\pm0.07$ & $\sim-0.45$              & $\sim-0.63$ & \ref{sptornado},\ref{sptornadocl} & 1,2 \\ 
G357.7+0.3\tablenotemark{d}            & 17:38:37.6 & --30:40:32 & 22           &                &                &                & $\sim0.04$ to $\sim-1.5$ &             & \ref{sptornado} & 1,3 \\
G359.1--0.5                            & 17:45:29.3 & --29:57:35 & 19          & $3.29\pm1.84$  & $6.81\pm2.46$ & $-1.30\pm1.19$ & $-1.9$ to $-0.5$           & $\sim-0.8$ & \ref{spg359.1-0.5},\ref{spg359.1-0.5cl}& 1,3,4 \\ 
G0.0+0.0 (Sgr A)\tablenotemark{d}      & 17:45:42.9 & --29:00:39 & $3\times4$   &                &                &                & $-0.44\pm0.02$           & $-0.42\pm0.02$ & \ref{spsgra},\ref{spsgracl} & 1,7,8 \\ 
G0.33+0.04\tablenotemark{d}            & 17:46:15   & --28:38:00 & $14\times8$  &                &                &                & $-2.0$ to $-1.5$         &             & \ref{speastarc}                & 9 \\
G0.9+0.1\tablenotemark{e}              & 17:47:22.5 & --28:09:19 & 3            & $5.46\pm0.35$  & $9.08\pm1.12$  & $-0.91\pm0.25$ & $-0.35$ to 0.08          & $-0.35$ & \ref{spg0.9+0.1},\ref{spg359.1-0.5cl} & 1,5 \\
G1.0--0.2\tablenotemark{e}             & 17:48:42.5 & --28:09:11 & 3$\times$4   & $8.52\pm0.53$  & $9.90\pm0.82$  & $-0.27\pm0.18$ & $-0.92$ to $-0.55$       & $-0.47$ & \ref{spg0.9+0.1},\ref{spg359.1-0.5cl} & 1,6 \\
\enddata
\tablenotetext{a}{The figure number featuring the SNR is listed here.}
\tablenotetext{b}{References that discuss object --- 1: \citet{g94}, 2: \citet{b85}, 3: \citet{r84}, 4: \citet{u92}, 5: \citet{h87}, 6: \citet{l92}, 7: \citet{p89}, 8: \citet{y87}, 9: \citet{ka96}}
\tablenotetext{c}{The Tornado is a candidate supernova remnant.}
\tablenotetext{d}{Source is only partially surveyed or confused with other sources, so no integrated characteristics are given.}
\tablenotetext{e}{Flux densities are given for the shell component only.}

\end{deluxetable}

\begin{deluxetable}{l|ccc|ccc}
\tablecaption{Flux Contributions by Thermal and Nonthermal Sources at 3.5 and 6 cm \label{fluxfrac}}
\tabletypesize{\scriptsize}
\tablewidth{0pt}
\tablehead{
               & \multicolumn{3}{|c|}{Extended Catalog} & \multicolumn{3}{|c|}{Extended and Compact Catalogs\tablenotemark{a}} \\
\hline
\colhead{Type} & \colhead{6 cm} & \colhead{3.5 cm} & \colhead{$\alpha_{CX}$} & \colhead{6 cm} & \colhead{3.5 cm} & \colhead{$\alpha_{CX}$} \\ 
\colhead{} & \colhead{(Jy)/(\%)} & \colhead{(Jy)/(\%)} & \colhead{} & \colhead{(Jy)/(\%)} & \colhead{(Jy)/(\%)} & \colhead{} \\ 
}
\startdata
thermal    &$147\pm43$ ($15\pm4$)  &$173\pm33$ ($24\pm5$) &$0.29\pm0.63$  &$202\pm48$ ($19\pm4$) &$220\pm33$ ($28\pm4$) &$0.15\pm0.50$ \\
nonthermal &$841\pm44$ ($85\pm4$)  &$552\pm19$ ($76\pm3$) &$-0.75\pm0.11$ &$862\pm46$ ($81\pm4$) &$563\pm19$ ($72\pm2$) &$-0.76\pm0.11$ \\
total      &$987\pm87$ (100)       &$725\pm52$ (100)      &$-0.55\pm0.20$ &$1063\pm93$ (100)     &$783\pm52$ (100)      &$-0.54\pm0.20$ \\
\enddata
\tablenotetext{a}{Includes all sources listed in Tables \ref{srcX}, \ref{srcC}, and \ref{diffsrc}.  See text in \S\ \ref{fluxfracsec} for details.}

\end{deluxetable}

\clearpage

\clearpage

\begin{figure}[tbp]
\includegraphics[bb=0 200 600 600,clip,width=6.5in]{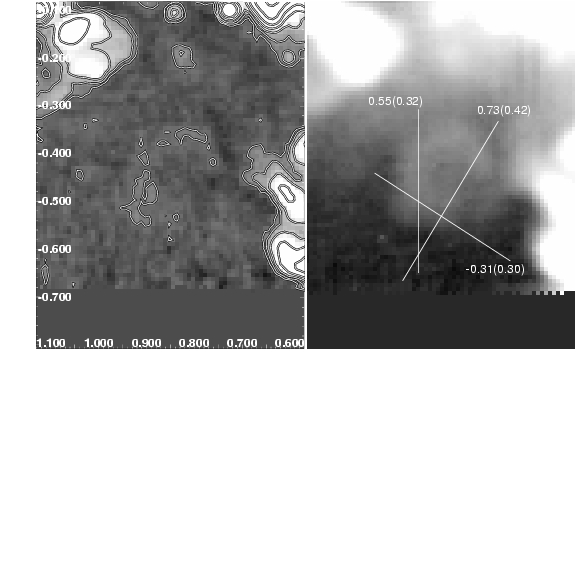}
\includegraphics[bb=0 200 600 600,clip,width=6.5in]{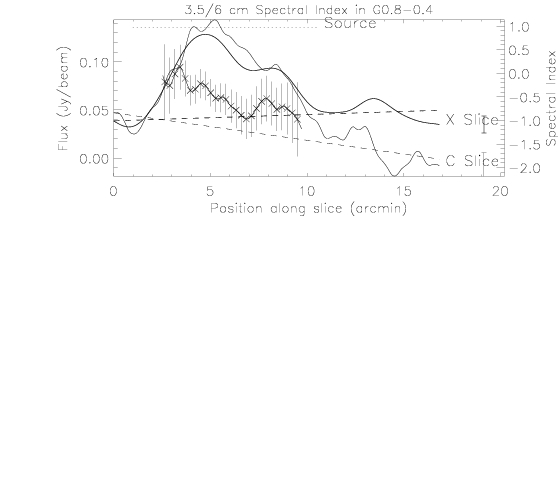}
\caption{Same as for Fig. \ref{sptornado}, but for the G0.8--0.4 complex.   The plotted slice values correspond to the slice with $\alpha_{CX}=-0.31\pm0.30$ with origin at the northeast. \label{spg0.8-0.4}}
\end{figure}

\end{document}